%% file: ewald.tex
\newcommand{\tchi}{\tilde{\chi}}

\documentstyle[epic,eepic,12pt]{aipproc}
\begin{document}
\bibliographystyle{aipproc}
\setcounter{footnote}{0}
\title{The Influence of Electrostatic Truncation on Simulations 
of Polarizable Systems\footnote{\normalsize The content of this publication does not necessarily reflect
the views or policies of the Department of Health and Human Services, nor
does mention of trade names, commercial products, or organization imply
endorsement by the U.S. Government.}}
\author{Steven W. Rick}
\address{Advanced Biomedical Computing Center, SAIC-Frederick,\\
NCI-Frederick Cancer Research and Development Center, \\
Frederick, Maryland, 27102}

\date{\ }

\maketitle

\begin{abstract}
Different schemes for the treatment of long-ranged electrostatic interactions
will be examined for water simulations using the polarizable
fluctuating charge potential.
Several different methods are compared, including Ewald sums,
potential shifting, spherical truncation and reaction field corrections.
For liquid water, properties such as the energy, pressure, dynamics and
structure are more sensitive to the treatment of the
long-ranged interactions with polarizable than with non-polarizable 
potentials.
\end{abstract}

\section*{Introduction}

Simulating systems with long-ranged electrostatic interactions using periodic
boundary conditions requires a treatment of the interactions beyond the
central simulation box, either using Ewald sums, truncation or modification
of the potential. The importance of the proper treatment of long-ranged
forces has been demonstrated for many systems, including pure 
water\cite{Watts,PangaliRaoBerne2,BrooksPettittKarplus,Prevost,Hunenberger}, hydrophobic aggregation\cite{NewBerne}, ionic solutions\cite{BrooksPettittKarplus}, and
proteins and peptides\cite{LoncharichBrooks,SmithPettitt,SchreiberSteinhauser,Darden,SteinbachBrooks}.
Nevertheless, many, if not most, simulations
of aqueous systems are done with periodic boundary conditions but
without using Ewald. There are two main
motivations for avoiding Ewald.
First, many seek to avoid the additional computational time 
that evaluating the Ewald 
sums requires, although algorithms such as the
cell multipole\cite{Greengard} and particle mesh 
Ewald\cite{EastwoodHockney,Darden2}
have made Ewald efficient for large systems.
Second, many seek to use periodic boundary conditions to avoid edge
effects but eliminate Ewald 
in the hope that 
having no direct interactions between periodic images
better represents infinite dilution for aqueous
solutes. As stated by Allen and Tildesley, Ewald methods ``will tend
to overemphasize the periodic nature of the model fluid\cite{Allen}.''
An alternative to Ewald and simple truncation are reaction field
methods in which the volume outside the cut-off distance is
treated as a dielectric continuum\cite{Allen,BarkerWatts}.
The influence of long-ranged interactions may be even more significant
for polarizable systems since the Coulomb or dipole-dipole interactions
will couple to the polarizable charges or dipoles. The recent widespread
use of polarizable water models, sometimes without using
Ewald\cite{Barnes,Ahlstrom,KozackJordon,vanBelle,BernardoLevy,Zhu,CaldwellKollman,Soetens,ChialvoCummings,DangChang,Siepmann},
suggests that an examination of the effects of truncation on
polarizable systems is necessary. The polarizable fluctuating charge
model for water will be used\cite{RickStuartBerne} to examine how
different truncation schemes and Ewald influence the structure and
dynamics of pure water.

Six different simulation methods will be used. 

\vspace{3. mm}

\noindent
{\bf A.} Ewald sums.

\vspace{2. mm}

\noindent
{\bf B.} Scaling by
the complementary error function, S$_B$(r)=erfc($\lambda$r).
This is simply the real space part of Ewald and $\lambda$ is set
equal to 5/L as is fairly standard\cite{Linse}.

\vspace{2. mm}

\noindent
{\bf C.} Scaling by erfc($\lambda$r), with the addition of
the Ewald self-term
and mean-field approximation for Fourier space term.

\vspace{2. mm}

\noindent
{\bf D.}  Scaling by

\begin{equation}
S_D(r) = \left\{ \begin{array}{ll}
       1 - 2 (r/r_{cut})^n + (r/r_{cut})^{2n} & \mbox{$r<r_{cut}$} \\
       0  & \mbox{$r>r_{cut}$} 
        \end{array}
\right.
\label{eq:sr1}
\end{equation}
Loncharich and Brooks used $n=2$ in 
Equation~\ref{eq:sr1}\cite{LoncharichBrooks}.
However, References~{\cite{BrooksPettittKarplus} and
~{\cite{Prevost} found that $n=1$ works better than $n=2$
for pure water. The present
simulations will use $n=1$ and $r_{cut}$=L/2, where L is the box
length.




\vspace{2. mm}

\noindent
{\bf E.} Nearest image truncation. Coulombic interactions between two-molecules
are included only if the
distance between center-of-masses is less than a cut-off distance, taken
to be L/2. 

\vspace{2. mm}

\noindent
{\bf F.} Reaction field correction to truncation.  The Coulomb interaction
becomes

\begin{equation}
E_{Coulomb}(r) = { 1 \over r} +
{\epsilon_{RF} -1 \over 2 \epsilon_{RF} +1 }
{r^2 \over r_{RF}^3}
- \left( {1 \over r_{cut}} + 
{\epsilon_{RF} -1 \over 2 \epsilon_{RF} +1 } {r_{cut}^2 \over r_{RF}^3}
\right)
\label{eq.erf}
\end{equation}
where $\epsilon_{RF}$ is the dielectric constant of the reaction field,
$r_{cut}$ is the cut-off radius and $r_{RF}$ is the radius of the reaction
field. H\"{u}nenberger and van Gunsteren have found that
$r_{RF}$=$r_{cut}$ is optimal\cite{Hunenberger} and so
the present simulations will use $r_{RF}$=$r_{cut}$=L/2 and
$\epsilon_{RF}$ is set equal to 79.
With $r_{RF}$=$r_{cut}$ and $\epsilon_{RF}$ much larger than 1,
the reaction field interaction becomes a scaling 
function acting on the Coulomb interaction equal to 1-(3/2)(r/r$_{cut}$)
+(1/2)(r/r$_{cut}$)$^3$.

\vspace{3. mm}

The different methods are plotted on Figure~\ref{fig:srs}, which
compares S(r)/r with the bare Coulomb interaction.
\begin{figure}[h]
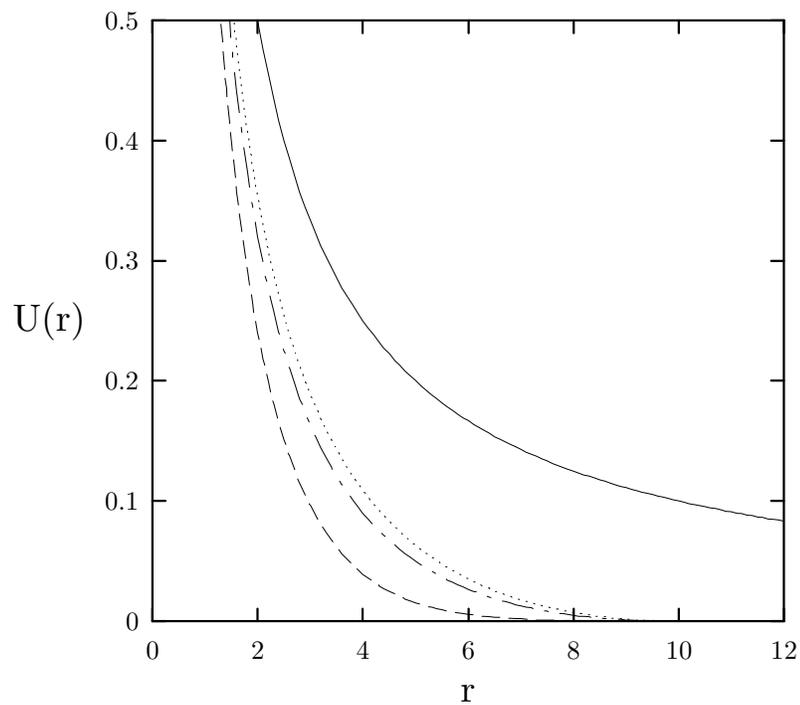

\include{cut3}
\caption{Comparison of the truncation methods with the Coulomb interaction,
1/r (solid line) showing erfc($\lambda$r)/r (dashed line), 
the shifted
potential S$_D$(r)/r (dot-dashed line) and the
reaction field interaction (dotted line).
}
\label{fig:srs}
\end{figure}
Notice that the functions S$_B$(r), S$_D$(r) and the reaction
field interaction are similar and modify the potential
over its whole range.
Notice also that at the cut-off, at 10, the Coulomb interaction is
0.1, only 20$\%$ lower than its value at hydrogen bonding distances
of 2 and is still far from zero.
Cut-off lengths are usually in the range of 8 to 10 {\AA}.

The periodic boundary conditions can either be applied to atoms
individually or by groups. If applied by groups (for example, by
molecules or residues), then the entire group is moved together
so that the nearest image between group centers (for example, the
center of mass) is used. If applied by atoms, then each atom is
moved individually. The second method can create large dipoles by 
splitting molecules across a box length. On the other hand, applying
the periodic boundary conditions to groups can suddenly shift
entire molecules across the box, resulting in discontinuities.
If the length scale of a molecule or residue 
(the distance from the group center to the outermost atom)
is {\bf R}
then the distance between two atoms can suddenly change from
$|${\bf r}+{\bf R}$|$ to $|${\bf r}-{\bf R}$|$.
These problems will not
be present if the potential equals zero at L/2-R.
For the methods in which the potential slowly approaches zero
at L/2 (methods {\bf A-D} and {\bf F}), atom based periodic boundary conditions
are used, since this was found to conserve energy better. (Although
the differences between atom and group based periodic boundary
conditions are not large for the Ewald simulation.)
For nearest image truncation with a spherical cut-off, the treatment
of nearest image makes a large difference and using atom-based
periodic boundary conditions leads to great instabilities in the
system, so a molecular based method was used.

The effects of the various treatments of the long-range forces on
the energy, pressure, dipole moment and pair correlation functions
will be examined. In addition, the dependence of dynamical
properties in terms of the translational and rotational 
diffusion constants will be examined. Convergence of these
properties will be examined for two different system sizes:
256 water molecules (corresponding to L=19.7{\AA} at a density of
1 g/cm$^3$) and 512 molecules (L=24.8{\AA}).

\section*{Potential Model}

The fluctuating charge (FQ) model is a polarizable potential model
in which the partial charges on atomic sites are treated as
variables which respond to changes in their 
environments\cite{RickStuartBerne}.
The model gives accurate predictions for liquid state properties. 
At T=298 and P=1 atm, the dielectric constant is 79.
The TIP4P-FQ model uses the geometry of the TIP4P water model
and includes Lennard-Jones interactions between oxygen sites and three
charge sites: two on the hydrogen atoms and one on the M-site 0.15{\AA} from
the oxygen atom\cite{JorgensenKlein}.
The FQ model has additional interactions between charge sites
on the same molecule. The energy is
\begin{equation}
E= E_{LJ} + E_{Coulomb} + E_{pol}
\label{eq:E1}
\end{equation}
where E$_{LJ}$ is the sum of all Lennard-Jones interactions 
between oxygen sites,
E$_{Coulomb}$ is the sum of all Coulomb interactions between
different molecules
\begin{equation}
E_{Coulomb}=\sum_{j<i}
\sum_{\alpha} \sum_{\beta}
{Q}_{i\alpha} {Q}_{j\beta}
/  {r}_{i\alpha,j\beta} 
\label{eq:ecoul}
\end{equation}
where Q$_{i\alpha}$ is the charge of atom $\alpha$ on molecule $i$
and E$_{pol}$ is the difference in the molecular energy between the
liquid and gas-phase,
\begin{equation}
E_{pol}=\sum_i \left( \sum_{\alpha} \tchi_{\alpha}^0 {Q}_{i\alpha}
+ {1\over 2} \sum_{\alpha} \sum_{\beta} {Q}_{i\alpha} {Q}_{i\beta}
{J}_{\alpha\beta}({r}_{i\alpha,i\beta}) - E_{gp} \right)
\label{eq:epol}
\end{equation}
where $E_{gp}$ is the gas-phase energy, $\tchi_{\alpha}^0$ is the Mulliken
electronegativity of the isolated atom and ${J}_{\alpha\beta}({r}_{i\alpha,i\beta})$
is the intramolecular interaction.
The gas-phase energy is the minimized energy of the isolated molecule.
With Ewald, E$_{Coulomb}$ becomes
\begin{equation}
E_{Coulomb}=\sum_{j<i}
\sum_{\alpha} \sum_{\beta}
{Q}_{i\alpha} {Q}_{j\beta} \: erfc(\lambda {r}_{i\alpha,j\beta})
/  {r}_{i\alpha,j\beta} 
\label{eq:ecoul2}
\end{equation}
and there are two additional energy terms: the Fourier space term
\begin{equation}
E_{FS} = {1 \over 2} { 4 \pi \over L^3}
\sum_{\bf G \ne 0} { 1 \over G^2} e^{-G^2/4 \lambda^2}
\left| \sum_j \sum_{\alpha} {Q}_{j\alpha} e^{i {\bf G} \cdot
{\bf r}_{j\alpha} } \right|^2
\label{eq:efs}
\end{equation}
and the self-term, which corrects for including $i=j$ terms in the
same box in E$_{FS}$,
\begin{equation}
E_{self}=-{1 \over 2} \sum_i \sum_{\alpha} \sum_{\beta}
{Q}_{i\alpha} {Q}_{i\beta}
\: erf(\lambda {r}_{i\alpha,i\beta})/{r}_{i\alpha,i\beta}.
\label{eq:eself}
\end{equation}
where erf(x) is the error function\cite{Allen}.
The screening parameter $\lambda$ was set equal to 5/L, 
256 lattice vectors were used in the Fourier space sum
and conducting boundary conditions were used.
There is a self energy for the reaction field method also,
which is given by\cite{Hunenberger}
\begin{equation}
E_{self}={1 \over 2} \sum_i \sum_{\alpha} \sum_{\beta}
{Q}_{i\alpha} {Q}_{i\beta} \:
{\epsilon_{RF} -1 \over 2 \epsilon_{RF} +1 }
{{r}_{i\alpha,i\beta}^2 \over r_{RF}^3 }.
\label{eq:eselfRF}
\end{equation}
Model {\bf C} consists of the Coulomb energy from
Equation~\ref{eq:ecoul2} and the self term (Equation~\ref{eq:eself})
plus an approximate Fourier space term
\begin{equation}
E_{FS} = 
{1 \over 2} \sum_i \sum_{\alpha} {Q}_{i\alpha} \langle \phi_{\alpha}^{FS}
\rangle
\label{eq:eselfRF2}
\end{equation}
with 
\begin{equation}
\langle \phi_{\alpha}^{FS} \rangle = \left\langle {1 \over N}
\sum_{\bf G \ne 0} { 1 \over G^2} e^{-G^2/4 \lambda^2}
\sum_i e^{-i {\bf G} \cdot {\bf r}_{i\alpha} }
\sum_j \sum_{\beta} {Q}_{j\beta} e^{i {\bf G} \cdot
{\bf r}_{j\beta} } \right\rangle 
\label{eq:eselfRF3}
\end{equation}
which was calculated from the Ewald simulation.
These terms are constants for rigid molecules but by coupling
to the charges they can influence the dynamics.
For 256 water molecules and L=19.7{\AA}, $\langle \phi_H^{FS} \rangle=$0.0125 kcal/mol/e
and $\langle \phi_O^{FS} \rangle=$-0.0280 kcal/mol/e.

The set of charges which minimize Equation~\ref{eq:E1} are the
ground state charges, subject to a charge neutrality constraint on
each molecule.
Rather that solving for the charges exactly at each time step,
the method treats them as dynamical variables, which are propagated
in an extended Lagrangian formalism
at a low temperature so as to remain near the potential energy
minimum\cite{RickStuartBerne}.
The extended Lagangian method introduces a new complication when
using cut-offs. If the cut-offs introduce discontinuities in the
potentials (as method {\bf E} does) then propagating the charge
degree-of-freedom becomes more difficult, because the forces
on the charges are discontinuous. For the case of method {\bf E}, 
in order to prevent the charge degrees-of-freedom from getting too
hot and drifting away from the potential energy minimum, the exact
set of charges was found every 1000 times steps. In all other cases,
the charge temperature remains under 10 Kelvin for a 100 ps
simulation. 
In addition, we found that the discontinuities in spherical truncation
caused the system to gradually heat up (at about 3 K/ps) so
these simulations were all done with a Nos\'{e}-Hoover temperature
bath with a mass for the Nos\'{e} variable equal to
2.0 kcal/mol psec$^2$\cite{Nose,Hoover}.
All other simulations were done at constant E,V,N, except for
equilibration (at T,V,N) and as noted.
The simulations were done with a 1 fs time step and used SHAKE to
enforce bond constraints\cite{RyckaertCiccottiBerendsen}.
The Lennard-Jones interactions were calculated only between the
nearest periodic images.
This too introduces some discontinuities into the energy and forces
and sometimes switching functions are used for the Lennard-Jones
interactions. However, for a box length of 20{\AA}, the
TIP4P-FQ Lennard-Jones interaction at half the box length is only
1x10$^{-3}$ kcal/mol.
The data presented in the next section 
is from four separate 100 ps runs.

\section*{Results}

\noindent
{\bf Energy.}
The energies for the different simulations are listed in
Table~\ref{tab:energy}. The
parentheses give 95$\%$ confidence intervals.
\begin{table}[b]
\caption{Total potential energy, divided by the number of molecules,
 and energy components 
for the various treatments of long-ranged electrostatics
for two different sized systems, in kcal/mol.}
\begin{tabular}{lccccccc}
 & {\bf number of} & {\bf E$_{\bf tot}$} & {\bf E$_{\bf LJ}$} & {\bf E$_{\bf Coulomb}$} &
{\bf E$_{\bf FS}$} & {\bf E$_{\bf self}$} & {\bf E$_{\bf pol}$} \\
 & {\bf molecules} &  &      &        &      &       &      \\
\hline
 A & 256 & -9.86(5) & 2.30(7) & -17.2(2) & 0.018(1)& -0.659(4) & 5.6(1) \\
   & 512 & -9.85(5) & 2.24(5) &-17.4(2) &0.009(1)  &-0.321(1)  &5.6(1)  \\
 B & 256 &-8.85(9)&1.59(6) &-14.6(2)  &      &       & 4.2(1) \\
   & 512 & -9.4(1)$\:\;$ &1.90(9) &-16.1(3) &      &       &4.9(1)\\
 C & 256 &-9.84(3) & 2.30(4) & -17.1(1) & 0.025(1) & -0.658(2) & 5.6(1) \\
 D & 256 & -9.09(4) & 1.59(3) &-15.0(1)&   &       & 4.4(1)     \\
   & 512 & -9.34(9) &1.78(8) &-15.8(3) &   &   & 4.7(1) \\
 E & 256 &-10.21(3)$\:\;$ &2.5(1)$\:\;$  &-18.9(3)  &      &       &6.2(1) \\
   & 512 &-10.20(3)$\:\;$ &2.49(2)  &-18.8(1)  &      &       &6.1(1) \\
 F & 256 & -9.92(8) & 2.20(9) & -17.7(3) & & -0.051(1) & 5.6(1) \\
   & 512 & -9.94(5) & 2.31(5) & -17.9(2) & & -0.026(1) & 5.7(1) \\
\multicolumn{2}{l}{experiment} & -9.9$^a$$\,\;\;\;\;$ &      &        &      &       &      \\
\end{tabular}
a. Reference~\cite{JorgensenKlein}
\label{tab:energy}
\end{table}
The Ewald results ({\bf A}) do not show a system size dependence for the
total energy. The individual contributions (E$_{Coulomb}$, 
E$_{FS}$ and E$_{self}$) should be system size dependent, since they
use a value of $\lambda$ dependent on the box size, but the sum of these
three terms should be size independent.
The largest component of the Ewald
terms is E$_{Coulomb}$. The other terms E$_{FS}$ and E$_{self}$ make 
much smaller contributions.
However, removing these terms, which would correspond to using a 
complementary error function shifting function (method {\bf B}),
gives a much different energy, which is higher by 1 kcal/mol
for the 256 molecule system. For the larger system with a longer
r$_{cut}$ the energy is closer to the Ewald result.
If the self-term and a mean-field estimate of E$_{FS}$ (see
Equations~\ref{eq:eselfRF2} and ~\ref{eq:eselfRF3})
are added back
in (model {\bf C}), the results are improved considerably.
These results are essentially indistinguishable from the Ewald
results.

The use of the shifting function (model {\bf D}) gives energies
similar to shifting by the complementary error function and the
energies show a strong dependence on cut-off length.
Other studies with non-polarizable water potentials
have also found that
using the shifted potential, S$_D$(r), leads to an increase in the
energy. For the SPC potential with
r$_{cut}$=9.3{\AA}, the energy is 0.4 kcal/mol higher\cite{Prevost}
and for the TIPS potential with r$_{cut}$=8.0{\AA}, the
energy is 0.6 kcal/mol higher\cite{BrooksPettittKarplus}.
For the polarizable model with a similar cut-off distance (for 256
molecules, r$_{cut}$=9.85{\AA}) the difference in the energy
is greater.
Nearest image truncation (model {\bf E}), on the other hand, 
overestimates the energy by almost a half a kcal/mol and does
not improve with system size.
For a non-polarizable models, the energy with spherical truncation also
does not show much of a dependence on cut-off 
length\cite{PangaliRaoBerne2,JorgensenMadura} and
overestimates the energy, but only by 0.1 kcal/mol\cite{Hunenberger}.
Once again the differences between Ewald and other treatments
is greater for the polarizable model.
Another study using spherical truncation with $r_{cut}$=10.5{\AA} 
and the SPC-FQ model also finds a lower energy (-11.5 kcal/mol)\cite{Siepmann}
than the Ewald result (-9.9 kcal/mol)\cite{RickStuartBerne}.
For the RPOL model of water, which treats polarizability using
point inducible dipoles, spherical truncation (with r$_{cut}$=9{\AA})
only slightly overestimates the energy\cite{SmithDang,Dang}.
The cut-offs apparently are more severe for the charge-charge (1/r)
interactions than for the dipole-dipole (1/r$^3$) interactions.
In fact, the cut-offs introduce less errors for the RPOL model
than for non-polarizable models.

The reaction field method (model {\bf F}) gives very good agreement
with the Ewald results for the energy. 
This agreement, and the improvement over the shifting function
(method {\bf D}), is remarkable considering the similarity of
the treatment of the Coulomb interaction (see Figure~\ref{fig:srs}).
The reaction field interaction (Equation~\ref{eq.erf}) is greater
than S$_D$(r)/r so the electrostatic interactions are 
stronger, which leads to the lower energy. In addition, there is
the self-term (Equation~\ref{eq:eselfRF}) which acts to slightly
increase the
magnitude of the charges and also lowers the energy.
For non-polarizable water potentials, the energy using the reaction
field method also agrees well with the energy using Ewald\cite{Hunenberger}.

\vspace{4. mm}

\noindent
{\bf Pressure.} The pressure is a balance of repulsive and
attractive forces and so is sensitive to the treatment of
electrostatic interactions (Table~\ref{tab:pressure}).
\begin{table}[t]
\caption{Total pressure and pressure components, in kbar.}
\begin{tabular}{lccccc}
 & {\bf number of} & {\bf P$_{\bf tot}$} & {\bf P$_{\bf LJ}$} & {\bf P$_{\bf Coulomb}$} &
{\bf P$_{\bf FS}$} \\
 & {\bf molecules} &       &      &       &      \\
\hline
 A & 256 & $\;\;$0.02(4) & 50.7(7) & -52.0(7) & -0.038(3) \\
   & 512 & $\;\;$0.04(5)& 50.2(5) &-51.6(5) &-0.015(2) \\
 B & 256 &0.5(2) &43.3(6) &-44.2(1) & \\
   & 512 &0.3(1) &46.7(9) &-47.8(9) &        \\
 C & 256 & $\;\;$0.15(6) & 50.7(4) & -51.9(4) & \\
 D & 256 & -1.5(1)$\;$& 43.2(3)&-46.2(4)&        \\
   & 512 & -1.3(1)$\;$&45.4(9) &-48.1(9)  &        \\
 E & 256 &-0.7(2)$\;$&53(1)$\;\;\,$  &-55.4(9)   &        \\
   & 512 &$\,$-0.77(6)&52.8(2) &-55.0(3)  &        \\
 F & 256 & -2.9(1)$\;$&49.6(9)&-53.9(9) & \\
   & 512 & -2.4(1)$\;$&51.0(5)&-54.7(6) & \\
\multicolumn{2}{l}{experiment} & 0.001$\:$ &   &   &      \\
\end{tabular}
\label{tab:pressure}
\end{table}
Listed are the total pressure, plus the contributions to the virial
from the different interactions. The components do not add up to
the total because there is the additional ideal gas part.
Only the Ewald method gives the correct pressure. The reaction field
method, which gave an accurate energy, does not do well for the
pressure.  The differences in the pressure between Ewald and other
methods are more substantial for the polarizable model than for
non-polarizable models\cite{Prevost,Hunenberger}.
Like the energy, the pressure with Ewald is size independent. A previous
simulation using Ewald with a non-polarizable potential found that
the energy and pressure shows no size
dependence for systems of 64 or more molecules\cite{Hummer}.

\vspace{4. mm}

\noindent
{\bf Dipole moment.} The dipole moment, which in all cases is enhanced
relative to the gas-phase value of 1.85 Debye, correlates very well
with the energy.
\begin{table}[b]
\caption{Dipole moment (Debye), translational diffusion constant (10$^{-9}$
m$^2$/s) and $\tau_{NMR}$ (ps).}
\begin{tabular}{lcccc}
 & {\bf number of} & {\bf dipole moment} & {\bf diffusion constant} & 
 ${\bf \tau_{NMR}}$ \\
& {\bf molecules} &       &      &  \\
\hline
 A & 256 & 2.62(1) & 2.0(3) & 2.2(2) \\
   & 512 & 2.62(1) & 2.1(3) & 2.0(1) \\
 B & 256 &2.51(1) &3.1(3) & 1.1(2)\\
   & 512 & 2.56(1) &2.7(4) &1.5(2) \\
 C & 256 &2.62(1) & 1.9(2) & 2.1(1) \\
 D & 256 & 2.53(1)&3.1(4)&1.2(1)\\
   & 512 & 2.55(1) &2.7(3) & 1.4(2)  \\
 E & 256 &2.66(1)&1.5(1)  &2.8(2)  \\
   & 512 &2.65(1) &1.8(1) &2.6(1) \\
 F & 256 &2.62(1) & 1.8(3) & 2.1(3) \\
   & 512 &2.62(1) & 1.9(1) & 2.2(1) \\
\multicolumn{2}{l}{experiment} & &2.3$^a$$\,\;\;$ &2.1$^b$$\,\;\;$    \\
\end{tabular}
a. Reference~\cite{Krynicki} \\
b. Reference~\cite{Jonas}
\label{tab:misc}
\end{table}
Treatments which give an accurate energies (models {\bf C} and {\bf F})
also have the same dipole moment as the Ewald method. Those with a
larger dipole moment (model {\bf E}) have a lower energy and those
with a smaller dipole moment (models {\bf B} and {\bf D}) have a
higher energy. 
It is the sensitivity of the charges that makes the
proper treatment of long-ranged electrostatics more important
for polarizable models.
The method {\bf C}, which did not work well, is improved considerably
just by adding the constant terms which couple to the charges to
give the right dipole moment. 

\vspace{4. mm}
\noindent
{\bf Dynamical properties.} Also listed in Table~\ref{tab:misc} are
the translational diffusion constant and the NMR relaxation time,
$\tau_{NMR}$, which
gives the time scale for rotations around the axis connecting
the hydrogen atoms\cite{RickStuartBerne,Impey}.
Methods which have an accurate energy and dipole moment ({\bf C} and
{\bf F}) have good transport properties. Methods with a higher
energy and a lower dipole moment ({\bf B} and {\bf D}) have
transport properties which are too fast. The diffusion constant
is larger and $\tau_{NMR}$ is smaller for these methods.
For the model ({\bf E}) with a lower energy and a larger dipole
moment, the transport properties are too slow.
For method {\bf E}, constant temperature dynamics is necessary to
avoid heating. Transport properties are sensitive to
how the velocity rescaling is done. It is preferable to use
constant E,V,N dynamics but the Nos\'{e}-Hoover method
for constant temperature dynamics can reproduce
diffusion properties well\cite{FrenkelSmit}.
Constant temperature dynamics with the Nos\'{e}-Hoover method
were run using Ewald as a check and the resulting
diffusion constant and $\tau_{NMR}$ were identical to the
constant E,V,N results. 
In other studies with non-polarizable water potentials, the
diffusion constant was found to be about the same as the
Ewald result using spherical truncation\cite{Hunenberger},
a reaction field\cite{Hunenberger} and
the shifting potential\cite{Prevost}.

\vspace{4. mm}

\noindent
{\bf Structure.} The radial distribution functions are sensitive to 
the treatment of the long-ranged interactions. Figure~\ref{fig:grD}
\begin{figure}[b]
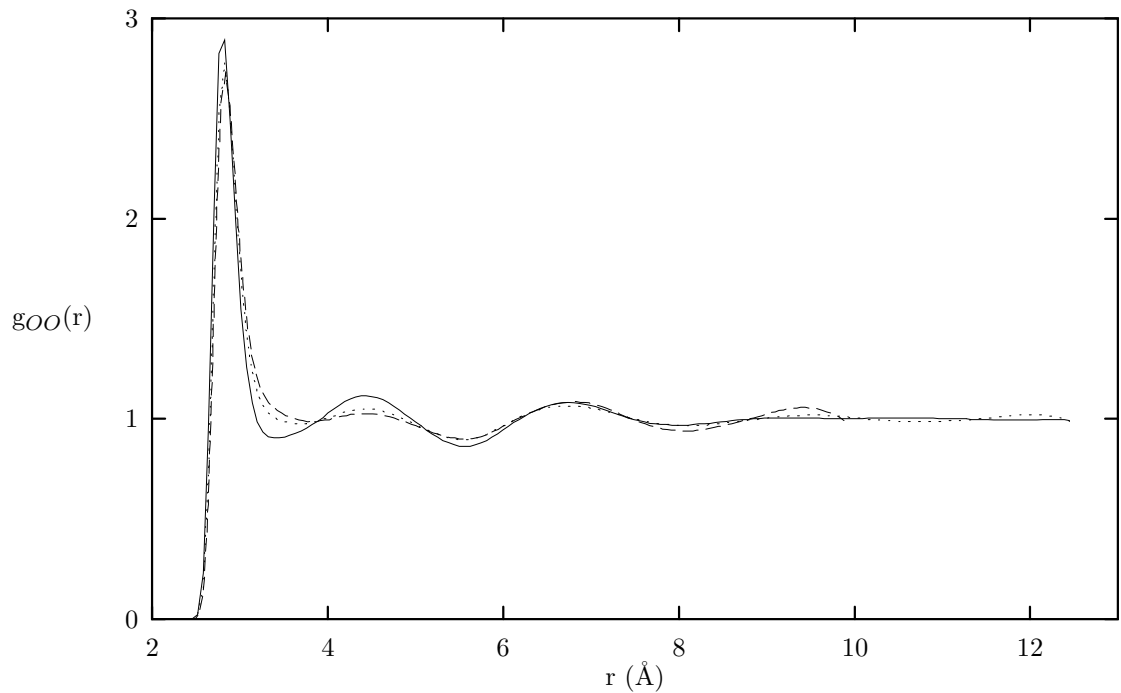

\include{grD}
\caption{Oxygen-oxygen radial distribution function with Ewald 
and 512 molecules (solid line), shifted potential, S$_D$(r),
 with 256 molecules
(dashed line), and shifted potential with 512 molecules (dotted line).}
\label{fig:grD}
\end{figure}
shows the oxygen-oxygen radial distribution function, g$_{OO}$(r), 
for Ewald with
512 molecules and the shifted potential (method {\bf D}) with
256 and 512 molecules. 
The Ewald g$_{OO}$(r)'s do not show a size dependence (data not shown)
but the shifted potential results do show a size dependence.
The first peaks do not have as much structure as Ewald, which is
consistent with the smaller charges that result using the shifted
potential. Also, there is structure around the cut-off distances
(9.85 {\AA} and 12.4 {\AA} for the different sized systems) due to
truncation effects. For non-polarizable models,
the agreement between method {\bf D} and Ewald is much 
better, although there still remains structure around
the cut-off distance\cite{BrooksPettittKarplus,Prevost}.
The results using method {\bf B} are similar to the method {\bf D}
results. They show a similar size dependence, but do not have
peaks at the cut-off distances. The interactions are sufficiently 
modified near the cut-off distance so that there are 
no truncation effects (see Figure~\ref{fig:srs}).

The results using the reaction field method agree well with the 
Ewald results, except for the peaks near the cut-off 
distance (Figure~\ref{fig:grRF}).
\begin{figure}[t]
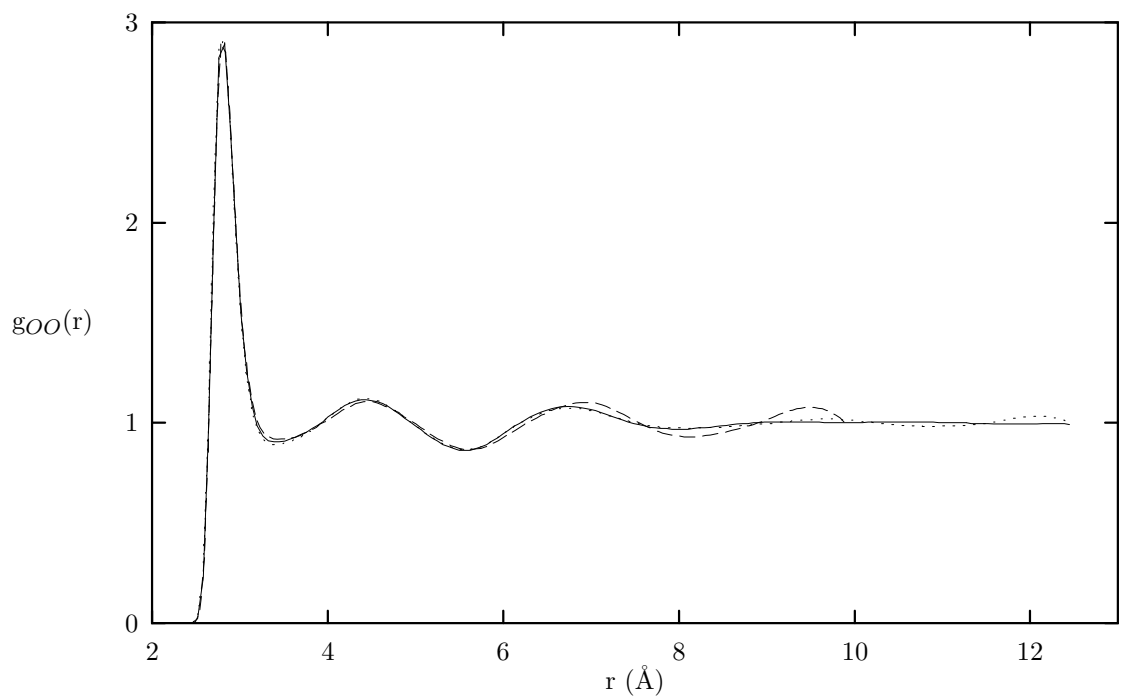

\include{grRF}
\caption{Oxygen-oxygen radial distribution function with Ewald 
and 512 molecules (solid line), reaction field method with 256 molecules
(dashed line), and reaction field method with 512 molecules (dotted line).}
\label{fig:grRF}
\end{figure}
The reaction field method also gives peaks at the cut-off
distance for non-polarizable potentials\cite{Hunenberger}.
With spherical truncation, the radial distribution functions are
similar to the Ewald results and are smooth at the cut-off
distances (Figure~\ref{fig:grE}).
\begin{figure}[t]
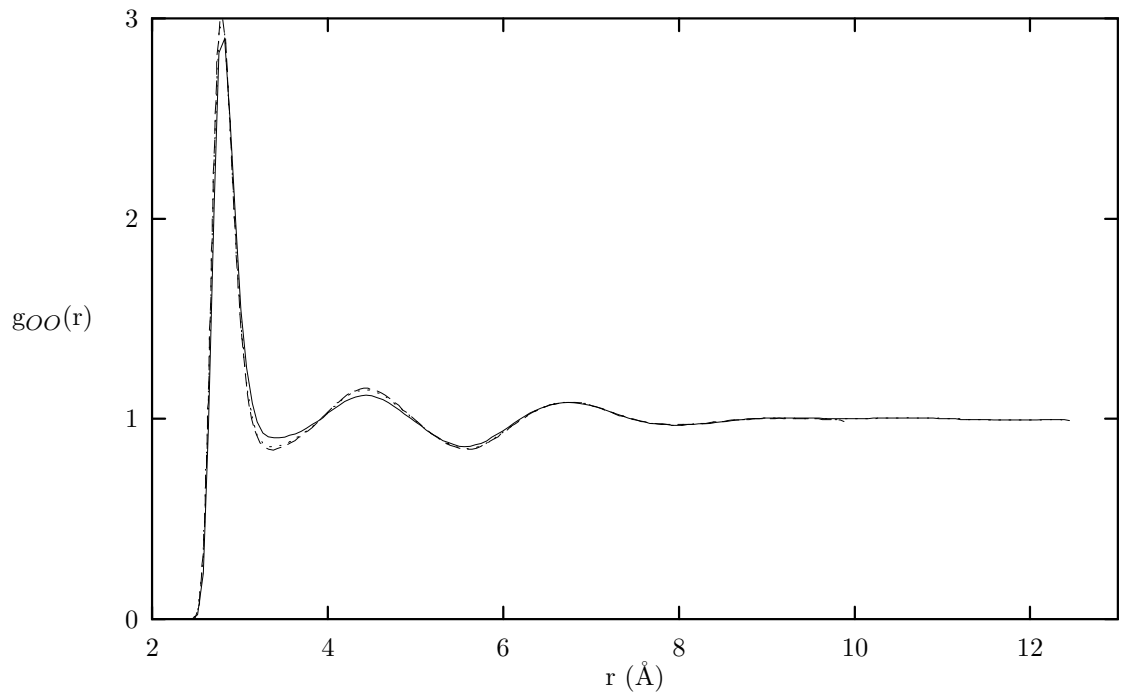

\include{grE}
\caption{Oxygen-oxygen radial distribution function with Ewald 
and 512 molecules (solid line), spherical truncation with 256 molecules
(dashed line), and spherical truncation with 512 molecules (dotted line).}
\label{fig:grE}
\end{figure}
There is slightly more structure in the first peaks consistent
with the larger charges.
There is no system size dependence in the
two spherical truncation results and the curves
are indistinguishable from
each other.
The g$_{OO}$(r) using method {\bf C} is identical to the Ewald
result and is not shown.

From the radial distribution functions, the pressure results can be
understood.
The simulation methods which produce the largest pressure difference
(models {\bf D} and {\bf E}) are also those which have peaks
in the g$_{OO}$(r) at the cut-off region.
These peaks will contribute to the virial and change the pressure.

\section*{Conclusions}

For systems with polarizable charges, the long-ranged interactions are
coupled to the charge distributions. Therefore, modifying
the treatment of these interactions will modify the charges and the Coulomb
interaction, q$_i$q$_j$/r, is changed not only by
changing 1/r to S(r)/r, but also by changing q$_i$ and q$_j$.
The results presented here indicate that the properties of
liquid water are more sensitive to the treatment of the
Coulomb interactions for polarizable systems than for
non-polarizable systems.
The reaction field correction
to truncation (method {\bf F}) is found to work well for the energy
and transport properties, although it does 
give a large negative pressure. 
The difference between the reaction field method and shifting the
potential by S$_D$(r) (Equation~\ref{eq:sr1}), which
did not work nearly as well, are small (see Figure~\ref{fig:srs}).
Subtle differences in the treatment of the electrostatics can cause
large differences in the results.
Another truncation method which does not work well, S(r)=erfc($\lambda$r),
can be made to give results almost identical to
the Ewald results just by adding two constant terms which couple
to the charges (compare methods {\bf B} and {\bf C}).
These terms represent the self-term of Ewald (Equation~\ref{eq:eself})
and a mean-field approximation to the Fourier-space term of Ewald
(Equations~\ref{eq:eselfRF2} and \ref{eq:eselfRF3}).
The fact that the mean-field approximation works so well suggests
that fluctuations in the Fourier-space and the forces from this
term are not important. 
Once these terms are added back into model {\bf B}, the charges
become equal to the charges with the Ewald method and all the
other properties including the energy, pressure, dynamics and
structure are accurately reproduced.

\vspace{0.5in}
\noindent
{\bf Acknowledgements}
This project has been funded in whole or in part with Federal funds from
the National Cancer Institute, National Institutes of Health, under
Contract No. NO1-CO-56000.

\bibliography{ewald}

\end{document}

%% file: cut3.tex
\setlength{\unitlength}{0.240900pt}
\begin{picture}(1275,1080)(-200,0)
\tenrm
\thicklines \path(220,113)(240,113)
\thicklines \path(1211,113)(1191,113)
\put(198,113){\makebox(0,0)[r]{0}}
\thicklines \path(220,302)(240,302)
\thicklines \path(1211,302)(1191,302)
\put(198,302){\makebox(0,0)[r]{0.1}}
\thicklines \path(220,491)(240,491)
\thicklines \path(1211,491)(1191,491)
\put(198,491){\makebox(0,0)[r]{0.2}}
\thicklines \path(220,679)(240,679)
\thicklines \path(1211,679)(1191,679)
\put(198,679){\makebox(0,0)[r]{0.3}}
\thicklines \path(220,868)(240,868)
\thicklines \path(1211,868)(1191,868)
\put(198,868){\makebox(0,0)[r]{0.4}}
\thicklines \path(220,1057)(240,1057)
\thicklines \path(1211,1057)(1191,1057)
\put(198,1057){\makebox(0,0)[r]{0.5}}
\thicklines \path(220,113)(220,133)
\thicklines \path(220,1057)(220,1037)
\put(220,68){\makebox(0,0){0}}
\thicklines \path(385,113)(385,133)
\thicklines \path(385,1057)(385,1037)
\put(385,68){\makebox(0,0){2}}
\thicklines \path(550,113)(550,133)
\thicklines \path(550,1057)(550,1037)
\put(550,68){\makebox(0,0){4}}
\thicklines \path(716,113)(716,133)
\thicklines \path(716,1057)(716,1037)
\put(716,68){\makebox(0,0){6}}
\thicklines \path(881,113)(881,133)
\thicklines \path(881,1057)(881,1037)
\put(881,68){\makebox(0,0){8}}
\thicklines \path(1046,113)(1046,133)
\thicklines \path(1046,1057)(1046,1037)
\put(1046,68){\makebox(0,0){10}}
\thicklines \path(1211,113)(1211,133)
\thicklines \path(1211,1057)(1211,1037)
\put(1211,68){\makebox(0,0){12}}
\thicklines \path(220,113)(1211,113)(1211,1057)(220,1057)(220,113)
\put(0,585){\makebox(0,0)[l]{\shortstack{\large U(r)}}}
\put(715,0){\makebox(0,0){\large r}}
\thinlines \path(385,1057)(385,1057)(393,1012)(402,971)(410,934)(418,900)(426,868)(435,839)(443,812)(451,787)(459,764)(468,742)(476,722)(484,703)(493,685)(501,668)(509,652)(517,637)(526,623)(534,610)(542,597)(550,585)(559,573)(567,563)(575,552)(583,542)(592,533)(600,523)(608,515)(616,506)(625,498)(633,491)(641,483)(649,476)(658,469)(666,463)(674,456)(682,450)(691,444)(699,438)(707,433)(716,428)(724,422)(732,418)(740,413)(749,408)(757,403)(765,399)(773,395)(782,391)(790,387)
\thinlines \path(790,387)(798,383)(806,379)(815,375)(823,372)(831,368)(839,365)(848,361)(856,358)(864,355)(872,352)(881,349)(889,346)(897,343)(905,341)(914,338)(922,335)(930,333)(938,330)(947,327)(955,325)(963,323)(972,320)(980,318)(988,316)(996,314)(1005,312)(1013,310)(1021,308)(1029,306)(1038,304)(1046,302)(1054,300)(1062,298)(1071,296)(1079,295)(1087,293)(1095,291)(1104,290)(1112,288)(1120,286)(1128,285)(1137,283)(1145,282)(1153,280)(1161,279)(1170,277)(1178,276)(1186,274)(1194,273)(1203,272)
\thinlines \path(1203,272)(1211,270)(1211,270)
\thinlines \path(348,1053)(348,1053)(348,1051)
\thinlines \path(349,1041)(349,1041)(349,1039)
\thinlines \path(351,1029)(351,1029)(351,1027)
\thinlines \path(352,1017)(352,1017)(352,1015)
\thinlines \path(353,1005)(353,1005)(353,1003)
\thinlines \path(355,992)(355,992)(355,990)
\thinlines \path(356,980)(356,980)(356,978)
\thinlines \path(358,968)(358,968)(358,966)
\thinlines \path(359,956)(359,956)(359,954)
\thinlines \path(361,944)(361,944)(361,942)
\thinlines \path(362,932)(362,932)(362,930)
\thinlines \path(364,919)(364,919)(364,917)
\thinlines \path(365,907)(365,907)(366,905)
\thinlines \path(367,895)(367,895)(367,893)
\thinlines \path(369,883)(369,883)(369,881)
\thinlines \path(371,871)(371,871)(371,869)
\thinlines \path(372,859)(372,859)(373,857)
\thinlines \path(374,847)(374,847)(375,845)
\thinlines \path(376,834)(376,834)(376,832)
\thinlines \path(378,822)(378,822)(378,820)
\thinlines \path(380,810)(380,810)(380,808)
\thinlines \path(382,798)(382,798)(382,796)
\thinlines \path(384,786)(384,786)(384,784)
\thinlines \path(386,774)(386,774)(387,772)
\thinlines \path(388,762)(388,762)(389,760)
\thinlines \path(391,750)(391,750)(391,748)
\thinlines \path(393,738)(393,738)(393,736)
\thinlines \path(395,726)(395,726)(396,724)
\thinlines \path(398,714)(398,714)(398,712)
\thinlines \path(400,702)(400,702)(401,700)
\thinlines \path(403,690)(403,690)(403,688)
\thinlines \path(405,678)(405,678)(406,676)
\thinlines \path(408,666)(408,666)(409,664)
\thinlines \path(411,654)(411,654)(411,652)
\thinlines \path(414,642)(414,642)(414,640)
\thinlines \path(417,630)(417,630)(417,628)
\thinlines \path(420,618)(420,618)(420,616)
\thinlines \path(423,606)(423,606)(423,604)
\thinlines \path(426,594)(426,594)(426,593)(427,592)
\thinlines \path(429,582)(429,582)(430,581)
\thinlines \path(433,571)(433,571)(433,569)
\thinlines \path(436,559)(436,559)(437,557)
\thinlines \path(440,547)(440,547)(441,545)
\thinlines \path(444,535)(444,535)(444,534)
\thinlines \path(448,524)(448,524)(448,522)
\thinlines \path(452,512)(452,512)(452,510)
\thinlines \path(456,501)(456,501)(456,499)
\thinlines \path(460,489)(460,489)(461,487)
\thinlines \path(464,478)(464,478)(465,476)
\thinlines \path(469,466)(469,466)(470,464)
\thinlines \path(474,455)(474,455)(475,453)
\thinlines \path(479,444)(479,444)(480,442)
\thinlines \path(484,432)(484,432)(484,431)(485,430)
\thinlines \path(489,421)(489,421)(490,419)
\thinlines \path(494,410)(494,410)(495,408)
\thinlines \path(500,399)(500,399)(501,398)(501,397)
\thinlines \path(506,388)(506,388)(507,387)
\thinlines \path(512,378)(512,378)(513,376)
\thinlines \path(518,367)(518,367)(519,365)
\thinlines \path(524,356)(524,356)(525,355)
\thinlines \path(531,346)(531,346)(532,344)
\thinlines \path(538,336)(538,336)(539,334)
\thinlines \path(545,326)(545,326)(546,324)
\thinlines \path(552,316)(552,316)(554,314)
\thinlines \path(560,306)(560,306)(561,304)
\thinlines \path(568,296)(568,296)(569,295)
\thinlines \path(576,287)(576,287)(577,285)
\thinlines \path(584,278)(584,278)(586,276)
\thinlines \path(593,269)(593,269)(594,267)
\thinlines \path(602,260)(602,260)(603,259)
\thinlines \path(611,251)(611,251)(612,250)
\thinlines \path(620,243)(620,243)(621,242)
\thinlines \path(629,235)(629,235)(631,234)
\thinlines \path(639,227)(639,227)(641,226)
\thinlines \path(649,220)(649,220)(649,220)(651,219)
\thinlines \path(659,213)(659,213)(661,212)
\thinlines \path(670,206)(670,206)(671,205)
\thinlines \path(680,199)(680,199)(682,198)
\thinlines \path(691,193)(691,193)(693,192)
\thinlines \path(702,187)(702,187)(704,186)
\thinlines \path(713,181)(713,181)(715,180)
\thinlines \path(724,175)(724,175)(726,175)
\thinlines \path(735,170)(735,170)(737,169)
\thinlines \path(747,165)(747,165)(749,165)(749,164)
\thinlines \path(758,160)(758,160)(760,160)
\thinlines \path(770,156)(770,156)(772,155)
\thinlines \path(782,152)(782,152)(784,151)
\thinlines \path(794,148)(794,148)(796,147)
\thinlines \path(806,144)(806,144)(806,144)(808,144)
\thinlines \path(818,141)(818,141)(820,140)
\thinlines \path(830,138)(830,138)(831,137)(832,137)
\thinlines \path(842,135)(842,135)(844,134)
\thinlines \path(854,132)(854,132)(856,132)(857,132)
\thinlines \path(867,129)(867,129)(869,129)
\thinlines \path(879,127)(879,127)(881,127)(881,127)
\thinlines \path(891,125)(891,125)(894,125)
\thinlines \path(904,123)(904,123)(905,123)(906,123)
\thinlines \path(916,121)(916,121)(918,121)
\thinlines \path(929,120)(929,120)(930,120)(931,120)
\thinlines \path(941,118)(941,118)(943,118)
\thinlines \path(954,117)(954,117)(955,117)(956,117)
\thinlines \path(966,116)(966,116)(968,116)
\thinlines \path(979,115)(979,115)(980,115)(981,115)
\thinlines \path(991,115)(991,115)(993,114)
\thinlines \path(1004,114)(1004,114)(1005,114)(1006,114)
\thinlines \path(1016,113)(1016,113)(1018,113)
\thinlines \path(1029,113)(1029,113)(1029,113)(1031,113)
\thinlines \path(1041,113)(1041,113)(1044,113)
\thinlines \path(1054,113)(1054,113)(1054,113)(1056,113)
\thinlines \path(1067,113)(1067,113)(1069,113)
\thinlines \path(1079,113)(1079,113)(1081,113)
\thinlines \path(1092,113)(1092,113)(1094,113)
\thinlines \path(1104,113)(1104,113)(1106,113)
\thinlines \path(1117,113)(1117,113)(1119,113)
\thinlines \path(1129,113)(1129,113)(1131,113)
\thinlines \path(1142,113)(1142,113)(1144,113)
\thinlines \path(1155,113)(1155,113)(1157,113)
\thinlines \path(1167,113)(1167,113)(1169,113)
\thinlines \path(1180,113)(1180,113)(1182,113)
\thinlines \path(1192,113)(1192,113)(1194,113)
\thinlines \path(1205,113)(1205,113)(1207,113)
\thinlines \path(327,1051)(327,1051)(330,1023)
\thinlines \path(331,1008)(331,1008)(333,980)
\thinlines \path(334,966)(334,966)(336,950)(337,937)
\thinlines \path(338,923)(338,923)(341,895)
\thinlines \path(342,881)(342,881)(344,863)(345,852)
\thinlines \path(347,838)(347,838)(350,810)
\thinlines \path(351,796)(351,796)(352,788)(355,767)
\thinlines \path(356,753)(356,753)(360,725)
\thinlines \path(362,711)(362,711)(366,683)
\thinlines \path(368,669)(368,669)(369,663)(372,641)
\thinlines \path(375,626)(375,626)(377,612)(379,598)
\thinlines \path(382,584)(382,584)(385,566)(387,556)
\thinlines \path(390,542)(390,542)(393,525)(396,515)
\thinlines \path(399,501)(399,501)(402,488)(405,473)
\thinlines \path(409,459)(409,459)(410,455)(416,432)
\thinlines \path(420,418)(420,418)(426,398)(429,391)
\thinlines \path(433,377)(433,377)(435,373)(443,351)
\thinlines \path(448,337)(448,337)(451,330)(459,312)(460,311)
\thinlines \path(466,298)(466,298)(468,295)(476,279)(479,273)
\thinlines \path(487,261)(487,261)(493,252)(501,240)(503,238)
\thinlines \path(512,226)(512,226)(517,219)(526,210)(531,205)
\thinlines \path(541,195)(541,195)(542,194)(550,187)(559,181)(563,178)
\thinlines \path(575,170)(575,170)(575,169)(583,164)(592,160)(600,156)
\thinlines \path(613,150)(613,150)(616,148)(625,145)(633,142)(639,140)
\thinlines \path(653,136)(653,136)(658,135)(666,133)(674,131)(681,129)
\thinlines \path(695,127)(695,127)(699,126)(707,125)(716,124)(723,123)
\thinlines \path(737,121)(737,121)(740,121)(749,120)(757,119)(765,119)(765,119)
\thinlines \path(779,118)(779,118)(782,118)(790,117)(798,117)(806,116)(808,116)
\thinlines \path(822,115)(822,115)(823,115)(831,115)(839,115)(848,115)(850,115)
\thinlines \path(865,114)(865,114)(872,114)(881,114)(889,114)(893,114)
\thinlines \path(907,114)(907,114)(914,114)(922,114)(930,114)(936,113)
\thinlines \path(950,113)(950,113)(955,113)(963,113)(972,113)(978,113)
\thinlines \path(993,113)(993,113)(996,113)(1005,113)(1013,113)(1021,113)(1021,113)
\thinlines \path(1035,113)(1035,113)(1038,113)(1046,113)(1054,113)(1062,113)(1064,113)
\thinlines \path(1078,113)(1078,113)(1079,113)(1087,113)(1095,113)(1104,113)(1106,113)
\thinlines \path(1121,113)(1121,113)(1128,113)(1137,113)(1145,113)(1149,113)
\thinlines \path(1163,113)(1163,113)(1170,113)(1178,113)(1186,113)(1192,113)
\thinlines \path(1206,113)(1206,113)(1211,113)(1211,113)
\thinlines \path(341,1057)(343,1030)
\thinlines \path(346,1007)(346,1007)(346,999)
\thinlines \path(349,975)(349,975)(352,946)(356,912)
\thinlines \path(359,889)(359,889)(360,881)
\thinlines \path(363,857)(363,857)(369,818)(372,795)
\thinlines \path(376,771)(376,771)(377,765)(377,763)
\thinlines \path(381,740)(381,740)(385,717)(393,678)
\thinlines \path(397,655)(397,655)(399,647)
\thinlines \path(404,624)(404,624)(410,600)(418,567)(419,564)
\thinlines \path(425,541)(425,541)(426,538)(428,534)
\thinlines \path(434,512)(434,512)(435,511)(443,486)(451,462)(455,454)
\thinlines \path(463,433)(463,433)(466,426)
\thinlines \path(475,405)(475,405)(476,403)(484,386)(493,370)(501,355)(502,352)
\thinlines \path(514,333)(514,333)(517,328)(518,327)
\thinlines \path(530,310)(530,310)(534,304)(542,293)(550,283)(559,273)(565,266)
\thinlines \path(580,251)(580,251)(583,248)(585,246)
\thinlines \path(600,233)(600,233)(600,233)(608,226)(616,219)(625,213)(633,207)(641,202)(642,201)
\thinlines \path(659,191)(659,191)(664,188)
\thinlines \path(681,179)(681,179)(682,178)(691,174)(699,170)(707,167)(716,163)(724,160)(728,159)
\thinlines \path(745,152)(745,152)(749,151)(751,150)
\thinlines \path(769,145)(769,145)(773,144)(782,142)(790,139)(798,137)(806,135)(815,134)(817,133)
\thinlines \path(835,129)(835,129)(839,129)(841,128)
\thinlines \path(859,125)(859,125)(864,125)(872,124)(881,122)(889,121)(897,121)(905,120)(908,119)
\thinlines \path(926,118)(926,118)(930,117)(932,117)
\thinlines \path(951,116)(951,116)(955,116)(963,115)(972,115)(980,114)(988,114)(996,114)(999,114)
\thinlines \path(1018,113)(1018,113)(1021,113)(1024,113)
\thinlines \path(1042,113)(1042,113)(1046,113)(1054,113)(1062,113)(1071,113)(1079,113)(1087,113)(1091,113)
\thinlines \path(1109,113)(1109,113)(1112,113)(1115,113)
\thinlines \path(1134,113)(1134,113)(1137,113)(1145,113)(1153,113)(1161,113)(1170,113)(1178,113)(1183,113)
\thinlines \path(1201,113)(1201,113)(1203,113)(1207,113)
\end{picture}

%% file: grD.tex
\setlength{\unitlength}{0.240900pt}
\begin{picture}(1800,1080)(0,0)
\tenrm
\thicklines \path(220,113)(240,113)
\thicklines \path(1736,113)(1716,113)
\put(198,113){\makebox(0,0)[r]{0}}
\thicklines \path(220,428)(240,428)
\thicklines \path(1736,428)(1716,428)
\put(198,428){\makebox(0,0)[r]{1}}
\thicklines \path(220,742)(240,742)
\thicklines \path(1736,742)(1716,742)
\put(198,742){\makebox(0,0)[r]{2}}
\thicklines \path(220,1057)(240,1057)
\thicklines \path(1736,1057)(1716,1057)
\put(198,1057){\makebox(0,0)[r]{3}}
\thicklines \path(220,113)(220,133)
\thicklines \path(220,1057)(220,1037)
\put(220,68){\makebox(0,0){2}}
\thicklines \path(496,113)(496,133)
\thicklines \path(496,1057)(496,1037)
\put(496,68){\makebox(0,0){4}}
\thicklines \path(771,113)(771,133)
\thicklines \path(771,1057)(771,1037)
\put(771,68){\makebox(0,0){6}}
\thicklines \path(1047,113)(1047,133)
\thicklines \path(1047,1057)(1047,1037)
\put(1047,68){\makebox(0,0){8}}
\thicklines \path(1323,113)(1323,133)
\thicklines \path(1323,1057)(1323,1037)
\put(1323,68){\makebox(0,0){10}}
\thicklines \path(1598,113)(1598,133)
\thicklines \path(1598,1057)(1598,1037)
\put(1598,68){\makebox(0,0){12}}
\thicklines \path(220,113)(1736,113)(1736,1057)(220,1057)(220,113)
\put(0,585){\makebox(0,0)[l]{\shortstack{g$_{OO}$(r)}}}
\put(978,23){\makebox(0,0){r ({\AA})}}
\thinlines \path(220,113)(223,113)(231,113)(240,113)(248,113)(257,113)(265,113)(274,113)(282,113)(291,120)(300,185)(308,414)(317,760)(325,1002)(334,1023)(342,898)(351,735)(359,600)(368,509)(377,453)(385,423)(394,407)(402,400)(411,398)(419,398)(428,399)(436,402)(445,405)(454,409)(462,413)(471,419)(479,424)(488,430)(496,437)(505,443)(514,449)(522,455)(531,459)(539,462)(548,464)(556,464)(565,463)(573,461)(582,458)(591,453)(599,448)(608,442)(616,435)(625,429)(633,423)(642,417)
\thinlines \path(642,417)(650,410)(659,404)(668,399)(676,394)(685,391)(693,388)(702,385)(710,385)(719,385)(727,387)(736,390)(745,394)(753,399)(762,404)(770,410)(779,416)(787,422)(796,427)(805,432)(813,437)(822,441)(830,445)(839,448)(847,450)(856,452)(864,453)(873,454)(882,453)(890,452)(899,451)(907,449)(916,447)(924,444)(933,441)(941,438)(950,435)(959,432)(967,429)(976,427)(984,425)(993,423)(1001,421)(1010,420)(1018,419)(1027,419)(1036,418)(1044,418)(1053,418)(1061,418)(1070,419)
\thinlines \path(1070,419)(1078,420)(1087,420)(1096,421)(1104,422)(1113,423)(1121,424)(1130,425)(1138,426)(1147,427)(1155,427)(1164,428)(1173,428)(1181,429)(1190,429)(1198,429)(1207,429)(1215,429)(1224,429)(1232,429)(1241,429)(1250,429)(1258,429)(1267,429)(1275,428)(1284,428)(1292,428)(1301,428)(1309,428)(1318,428)(1327,428)(1335,428)(1344,429)(1352,429)(1361,429)(1369,429)(1378,429)(1387,429)(1395,429)(1404,429)(1412,429)(1421,429)(1429,429)(1438,429)(1446,429)(1455,428)(1464,428)(1472,428)(1481,428)(1489,428)(1498,428)
\thinlines \path(1498,428)(1506,427)(1515,427)(1523,427)(1532,427)(1541,426)(1549,426)(1558,426)(1566,426)(1575,426)(1583,426)(1592,426)(1600,426)(1609,426)(1618,427)(1626,427)(1635,427)(1643,427)(1652,427)(1660,425)
\thinlines \path(220,113)(223,113)
\thinlines \path(235,113)(235,113)(240,113)(247,113)(253,113)(259,113)
\thinlines \path(271,113)(271,113)(274,113)(281,113)(287,113)(292,119)
\thinlines \path(295,128)(295,128)(299,149)
\thinlines \path(300,159)(300,159)(301,164)(302,179)
\thinlines \path(302,189)(302,189)(303,210)
\thinlines \path(304,220)(304,220)(305,240)
\thinlines \path(306,250)(306,250)(307,271)
\thinlines \path(307,281)(307,281)(308,293)(308,301)
\thinlines \path(308,312)(308,312)(309,332)
\thinlines \path(309,342)(309,342)(310,363)
\thinlines \path(310,373)(310,373)(311,393)
\thinlines \path(311,403)(311,403)(312,424)
\thinlines \path(312,434)(312,434)(313,455)
\thinlines \path(313,465)(313,465)(314,485)
\thinlines \path(314,495)(314,495)(315,516)
\thinlines \path(315,526)(315,526)(315,546)
\thinlines \path(316,557)(316,557)(316,577)
\thinlines \path(317,587)(317,587)(317,608)
\thinlines \path(317,618)(317,618)(318,638)
\thinlines \path(318,648)(318,648)(319,669)
\thinlines \path(319,679)(319,679)(320,700)
\thinlines \path(320,710)(320,710)(321,730)
\thinlines \path(321,740)(321,740)(321,761)
\thinlines \path(322,771)(322,771)(323,791)
\thinlines \path(323,802)(323,802)(324,822)
\thinlines \path(324,832)(324,832)(325,853)
\thinlines \path(326,863)(326,863)(326,883)
\thinlines \path(327,893)(327,893)(328,914)
\thinlines \path(328,924)(328,924)(328,924)(331,944)
\thinlines \path(333,954)(333,954)(335,970)(336,966)
\thinlines \path(337,956)(337,956)(340,935)
\thinlines \path(341,925)(341,925)(342,920)(343,905)
\thinlines \path(344,895)(344,895)(345,874)
\thinlines \path(346,864)(346,864)(347,844)
\thinlines \path(348,833)(348,833)(349,828)(350,813)
\thinlines \path(350,803)(350,803)(352,782)
\thinlines \path(352,772)(352,772)(354,752)
\thinlines \path(354,742)(354,742)(355,728)(356,721)
\thinlines \path(357,711)(357,711)(358,691)
\thinlines \path(359,681)(359,681)(361,660)
\thinlines \path(362,650)(362,650)(362,643)(363,630)
\thinlines \path(365,619)(365,619)(367,599)
\thinlines \path(368,589)(368,589)(369,576)(370,569)
\thinlines \path(372,559)(372,559)(375,538)
\thinlines \path(376,528)(376,528)(381,508)
\thinlines \path(383,498)(383,498)(389,478)(389,478)
\thinlines \path(394,469)(394,469)(396,464)(403,453)(405,451)
\thinlines \path(413,444)(413,444)(416,441)(423,437)(430,433)(433,432)
\thinlines \path(444,428)(444,428)(450,426)(457,425)(464,423)(467,424)
\thinlines \path(479,424)(479,424)(484,425)(491,426)(498,426)(502,428)
\thinlines \path(514,431)(514,431)(518,432)(525,433)(532,435)(537,435)
\thinlines \path(549,436)(549,436)(552,436)(559,436)(566,436)(573,435)(573,435)
\thinlines \path(585,433)(585,433)(586,433)(593,431)(600,429)(607,426)(607,426)
\thinlines \path(618,423)(618,423)(620,422)(627,419)(634,417)(640,414)
\thinlines \path(651,410)(651,410)(654,409)(661,406)(668,404)(673,402)
\thinlines \path(684,399)(684,399)(688,398)(695,398)(702,396)(708,396)
\thinlines \path(720,397)(720,397)(722,397)(729,398)(736,400)(743,402)(743,402)
\thinlines \path(753,407)(753,407)(756,409)(763,411)(770,416)(774,418)
\thinlines \path(784,423)(784,423)(790,427)(797,430)(804,433)(805,433)
\thinlines \path(815,438)(815,438)(817,439)(824,442)(831,444)(837,447)
\thinlines \path(848,450)(848,450)(851,451)(858,452)(865,453)(872,454)(872,454)
\thinlines \path(884,455)(884,455)(885,455)(892,454)(899,454)(906,453)(907,452)
\thinlines \path(919,449)(919,449)(919,449)(926,447)(933,445)(940,442)(941,442)
\thinlines \path(952,437)(952,437)(953,436)(960,433)(967,430)(973,428)
\thinlines \path(984,423)(984,423)(987,422)(994,420)(1001,417)(1006,416)
\thinlines \path(1018,413)(1018,413)(1021,412)(1028,411)(1035,410)(1041,410)
\thinlines \path(1053,409)(1053,409)(1055,409)(1062,409)(1069,409)(1075,410)(1077,410)
\thinlines \path(1089,412)(1089,412)(1089,412)(1096,413)(1103,415)(1109,416)(1112,416)
\thinlines \path(1124,419)(1124,419)(1130,420)(1137,422)(1143,424)(1147,425)
\thinlines \path(1158,428)(1158,428)(1164,429)(1171,431)(1177,433)(1181,434)
\thinlines \path(1192,438)(1192,438)(1198,439)(1204,441)(1211,442)(1215,443)
\thinlines \path(1227,445)(1227,445)(1232,445)(1238,446)(1245,446)(1251,445)
\thinlines \path(1263,444)(1263,444)(1266,443)(1272,441)(1279,439)(1285,437)
\thinlines \path(1296,432)(1296,432)(1300,430)(1306,425)
\thinlines \path(220,113)(221,113)
\thinlines \path(235,113)(235,113)(237,113)
\thinlines \path(251,113)(251,113)(253,113)
\thinlines \path(266,113)(266,113)(269,113)
\thinlines \path(282,113)(282,113)(282,113)(285,114)
\thinlines \path(292,122)(292,122)(292,124)
\thinlines \path(295,135)(295,135)(295,138)
\thinlines \path(297,149)(297,149)(298,151)
\thinlines \path(300,162)(300,162)(300,164)
\thinlines \path(300,176)(300,176)(300,178)
\thinlines \path(301,189)(301,189)(301,191)
\thinlines \path(302,203)(302,203)(302,205)
\thinlines \path(302,216)(302,216)(302,218)
\thinlines \path(303,230)(303,230)(303,232)
\thinlines \path(304,243)(304,243)(304,245)
\thinlines \path(304,257)(304,257)(304,259)
\thinlines \path(305,270)(305,270)(305,272)
\thinlines \path(305,284)(305,284)(306,286)
\thinlines \path(306,297)(306,297)(306,300)
\thinlines \path(307,311)(307,311)(307,313)
\thinlines \path(307,324)(307,324)(308,327)
\thinlines \path(308,338)(308,338)(308,338)(308,340)
\thinlines \path(308,351)(308,351)(309,354)
\thinlines \path(309,365)(309,365)(309,367)
\thinlines \path(309,378)(309,378)(309,381)
\thinlines \path(310,392)(310,392)(310,394)
\thinlines \path(310,406)(310,406)(310,408)
\thinlines \path(310,419)(310,419)(310,421)
\thinlines \path(311,433)(311,433)(311,435)
\thinlines \path(311,446)(311,446)(311,448)
\thinlines \path(311,460)(311,460)(312,462)
\thinlines \path(312,473)(312,473)(312,475)
\thinlines \path(312,487)(312,487)(312,489)
\thinlines \path(313,500)(313,500)(313,503)
\thinlines \path(313,514)(313,514)(313,516)
\thinlines \path(313,527)(313,527)(313,530)
\thinlines \path(314,541)(314,541)(314,543)
\thinlines \path(314,554)(314,554)(314,557)
\thinlines \path(314,568)(314,568)(314,570)
\thinlines \path(315,581)(315,581)(315,584)
\thinlines \path(315,595)(315,595)(315,597)
\thinlines \path(316,609)(316,609)(316,611)
\thinlines \path(316,622)(316,622)(316,624)
\thinlines \path(316,636)(316,636)(316,638)
\thinlines \path(317,649)(317,649)(317,651)
\thinlines \path(317,663)(317,663)(317,665)
\thinlines \path(318,676)(318,676)(318,678)
\thinlines \path(318,690)(318,690)(318,692)
\thinlines \path(318,703)(318,703)(319,706)
\thinlines \path(319,717)(319,717)(319,719)
\thinlines \path(319,730)(319,730)(319,733)
\thinlines \path(320,744)(320,744)(320,746)
\thinlines \path(320,757)(320,757)(320,760)
\thinlines \path(321,771)(321,771)(321,773)
\thinlines \path(321,784)(321,784)(321,787)
\thinlines \path(322,798)(322,798)(322,800)
\thinlines \path(322,812)(322,812)(322,814)
\thinlines \path(322,825)(322,825)(322,827)
\thinlines \path(323,839)(323,839)(323,841)
\thinlines \path(323,852)(323,852)(323,854)
\thinlines \path(324,866)(324,866)(324,868)
\thinlines \path(324,879)(324,879)(324,881)
\thinlines \path(325,893)(325,893)(325,895)
\thinlines \path(325,906)(325,906)(325,909)
\thinlines \path(326,920)(326,920)(326,922)
\thinlines \path(328,933)(328,933)(328,936)
\thinlines \path(329,947)(329,947)(330,949)
\thinlines \path(331,960)(331,960)(331,962)
\thinlines \path(332,974)(332,974)(333,976)
\thinlines \path(334,987)(334,987)(334,985)
\thinlines \path(335,973)(335,973)(336,971)
\thinlines \path(337,960)(337,960)(337,958)
\thinlines \path(338,946)(338,946)(339,944)
\thinlines \path(340,933)(340,933)(340,931)
\thinlines \path(341,920)(341,920)(341,917)
\thinlines \path(343,906)(343,906)(343,904)
\thinlines \path(343,892)(343,892)(344,890)
\thinlines \path(344,879)(344,879)(344,877)
\thinlines \path(345,865)(345,865)(345,863)
\thinlines \path(346,852)(346,852)(346,850)
\thinlines \path(347,838)(347,838)(347,836)
\thinlines \path(348,825)(348,825)(348,823)
\thinlines \path(349,811)(349,811)(349,809)
\thinlines \path(349,798)(349,798)(350,796)
\thinlines \path(350,784)(350,784)(350,782)
\thinlines \path(351,771)(351,771)(351,769)
\thinlines \path(352,757)(352,757)(352,755)
\thinlines \path(353,744)(353,744)(353,742)
\thinlines \path(354,730)(354,730)(354,728)
\thinlines \path(355,717)(355,717)(355,715)
\thinlines \path(356,703)(356,703)(356,701)
\thinlines \path(357,690)(357,690)(357,687)
\thinlines \path(358,676)(358,676)(358,674)
\thinlines \path(359,663)(359,663)(359,660)
\thinlines \path(360,649)(360,649)(360,647)
\thinlines \path(361,636)(361,636)(361,633)
\thinlines \path(362,622)(362,622)(362,620)
\thinlines \path(363,609)(363,609)(364,606)
\thinlines \path(365,595)(365,595)(365,593)
\thinlines \path(366,582)(366,582)(366,579)
\thinlines \path(367,568)(367,568)(368,566)
\thinlines \path(369,555)(369,555)(369,553)
\thinlines \path(371,541)(371,541)(371,539)
\thinlines \path(373,528)(373,528)(373,526)
\thinlines \path(375,514)(375,514)(375,512)
\thinlines \path(377,501)(377,501)(378,499)
\thinlines \path(380,488)(380,488)(381,486)
\thinlines \path(384,475)(384,475)(384,472)
\thinlines \path(388,462)(388,462)(389,459)
\thinlines \path(394,449)(394,449)(394,449)(395,447)
\thinlines \path(402,437)(402,437)(402,437)(404,436)
\thinlines \path(414,429)(414,429)(416,427)
\thinlines \path(428,423)(428,423)(431,423)
\thinlines \path(444,420)(444,420)(445,420)(447,420)
\thinlines \path(460,420)(460,420)(462,420)(462,421)
\thinlines \path(475,423)(475,423)(478,424)
\thinlines \path(490,428)(490,428)(493,428)
\thinlines \path(505,432)(505,432)(508,433)
\thinlines \path(520,438)(520,438)(522,439)(522,439)
\thinlines \path(535,441)(535,441)(538,442)
\thinlines \path(551,443)(551,443)(553,443)
\thinlines \path(566,443)(566,443)(569,443)
\thinlines \path(582,440)(582,440)(584,439)
\thinlines \path(597,435)(597,435)(599,434)
\thinlines \path(611,429)(611,429)(613,428)
\thinlines \path(625,423)(625,423)(627,422)
\thinlines \path(639,416)(639,416)(641,415)
\thinlines \path(653,410)(653,410)(655,409)
\thinlines \path(667,404)(667,404)(668,404)(670,403)
\thinlines \path(682,399)(682,399)(684,398)
\thinlines \path(697,396)(697,396)(700,395)
\thinlines \path(713,395)(713,395)(716,395)
\thinlines \path(729,398)(729,398)(731,398)
\thinlines \path(743,403)(743,403)(745,403)(746,404)
\thinlines \path(757,410)(757,410)(759,411)
\thinlines \path(771,417)(771,417)(773,418)
\thinlines \path(784,424)(784,424)(786,425)
\thinlines \path(797,431)(797,431)(800,432)
\thinlines \path(812,437)(812,437)(813,437)(814,438)
\thinlines \path(827,442)(827,442)(829,443)
\thinlines \path(842,446)(842,446)(844,446)
\thinlines \path(858,448)(858,448)(860,448)
\thinlines \path(873,448)(873,448)(876,448)
\thinlines \path(889,448)(889,448)(890,448)(892,447)
\thinlines \path(905,446)(905,446)(907,445)(908,445)
\thinlines \path(920,443)(920,443)(923,442)
\thinlines \path(936,439)(936,439)(938,438)
\thinlines \path(951,435)(951,435)(953,434)
\thinlines \path(966,431)(966,431)(967,430)(969,430)
\thinlines \path(981,427)(981,427)(984,427)
\thinlines \path(997,424)(997,424)(999,423)
\thinlines \path(1012,421)(1012,421)(1015,420)
\thinlines \path(1028,419)(1028,419)(1031,419)
\thinlines \path(1044,418)(1044,418)(1044,418)(1046,418)
\thinlines \path(1060,417)(1060,417)(1061,417)(1062,417)
\thinlines \path(1075,418)(1075,418)(1078,418)
\thinlines \path(1091,418)(1091,418)(1094,419)
\thinlines \path(1107,420)(1107,420)(1110,420)
\thinlines \path(1123,422)(1123,422)(1125,422)
\thinlines \path(1139,424)(1139,424)(1141,424)
\thinlines \path(1154,425)(1154,425)(1155,426)(1157,426)
\thinlines \path(1170,427)(1170,427)(1173,428)(1173,428)
\thinlines \path(1186,429)(1186,429)(1188,430)
\thinlines \path(1202,431)(1202,431)(1204,431)
\thinlines \path(1217,432)(1217,432)(1220,433)
\thinlines \path(1233,433)(1233,433)(1236,433)
\thinlines \path(1249,434)(1249,434)(1250,434)(1252,434)
\thinlines \path(1265,434)(1265,434)(1267,434)(1268,434)
\thinlines \path(1281,433)(1281,433)(1283,433)
\thinlines \path(1297,432)(1297,432)(1299,432)
\thinlines \path(1312,431)(1312,431)(1315,430)
\thinlines \path(1328,429)(1328,429)(1331,429)
\thinlines \path(1344,427)(1344,427)(1347,427)
\thinlines \path(1360,426)(1360,426)(1361,426)(1362,426)
\thinlines \path(1376,425)(1376,425)(1378,425)(1378,425)
\thinlines \path(1392,424)(1392,424)(1394,424)
\thinlines \path(1407,424)(1407,424)(1410,424)
\thinlines \path(1423,424)(1423,424)(1426,424)
\thinlines \path(1439,424)(1439,424)(1442,424)
\thinlines \path(1455,424)(1455,424)(1458,424)
\thinlines \path(1471,425)(1471,425)(1472,425)(1474,425)
\thinlines \path(1487,425)(1487,425)(1489,425)(1489,425)
\thinlines \path(1503,426)(1503,426)(1505,426)
\thinlines \path(1518,428)(1518,428)(1521,428)
\thinlines \path(1534,429)(1534,429)(1537,429)
\thinlines \path(1550,431)(1550,431)(1553,431)
\thinlines \path(1566,433)(1566,433)(1566,433)(1568,433)
\thinlines \path(1582,434)(1582,434)(1583,434)(1584,434)
\thinlines \path(1597,434)(1597,434)(1600,434)
\thinlines \path(1613,434)(1613,434)(1616,434)
\thinlines \path(1629,432)(1629,432)(1632,432)
\thinlines \path(1645,430)(1645,430)(1647,429)
\thinlines \path(1659,425)(1659,425)(1660,424)
\end{picture}

%% file: grRF.tex
\setlength{\unitlength}{0.240900pt}
\begin{picture}(1800,1080)(0,0)
\tenrm
\thicklines \path(220,113)(240,113)
\thicklines \path(1736,113)(1716,113)
\put(198,113){\makebox(0,0)[r]{0}}
\thicklines \path(220,428)(240,428)
\thicklines \path(1736,428)(1716,428)
\put(198,428){\makebox(0,0)[r]{1}}
\thicklines \path(220,742)(240,742)
\thicklines \path(1736,742)(1716,742)
\put(198,742){\makebox(0,0)[r]{2}}
\thicklines \path(220,1057)(240,1057)
\thicklines \path(1736,1057)(1716,1057)
\put(198,1057){\makebox(0,0)[r]{3}}
\thicklines \path(220,113)(220,133)
\thicklines \path(220,1057)(220,1037)
\put(220,68){\makebox(0,0){2}}
\thicklines \path(496,113)(496,133)
\thicklines \path(496,1057)(496,1037)
\put(496,68){\makebox(0,0){4}}
\thicklines \path(771,113)(771,133)
\thicklines \path(771,1057)(771,1037)
\put(771,68){\makebox(0,0){6}}
\thicklines \path(1047,113)(1047,133)
\thicklines \path(1047,1057)(1047,1037)
\put(1047,68){\makebox(0,0){8}}
\thicklines \path(1323,113)(1323,133)
\thicklines \path(1323,1057)(1323,1037)
\put(1323,68){\makebox(0,0){10}}
\thicklines \path(1598,113)(1598,133)
\thicklines \path(1598,1057)(1598,1037)
\put(1598,68){\makebox(0,0){12}}
\thicklines \path(220,113)(1736,113)(1736,1057)(220,1057)(220,113)
\put(0,585){\makebox(0,0)[l]{\shortstack{g$_{OO}$(r)}}}
\put(978,23){\makebox(0,0){r ({\AA})}}
\thinlines \path(220,113)(223,113)(231,113)(240,113)(248,113)(257,113)(265,113)(274,113)(282,113)(291,120)(300,185)(308,414)(317,760)(325,1002)(334,1023)(342,898)(351,735)(359,600)(368,509)(377,453)(385,423)(394,407)(402,400)(411,398)(419,398)(428,399)(436,402)(445,405)(454,409)(462,413)(471,419)(479,424)(488,430)(496,437)(505,443)(514,449)(522,455)(531,459)(539,462)(548,464)(556,464)(565,463)(573,461)(582,458)(591,453)(599,448)(608,442)(616,435)(625,429)(633,423)(642,417)
\thinlines \path(642,417)(650,410)(659,404)(668,399)(676,394)(685,391)(693,388)(702,385)(710,385)(719,385)(727,387)(736,390)(745,394)(753,399)(762,404)(770,410)(779,416)(787,422)(796,427)(805,432)(813,437)(822,441)(830,445)(839,448)(847,450)(856,452)(864,453)(873,454)(882,453)(890,452)(899,451)(907,449)(916,447)(924,444)(933,441)(941,438)(950,435)(959,432)(967,429)(976,427)(984,425)(993,423)(1001,421)(1010,420)(1018,419)(1027,419)(1036,418)(1044,418)(1053,418)(1061,418)(1070,419)
\thinlines \path(1070,419)(1078,420)(1087,420)(1096,421)(1104,422)(1113,423)(1121,424)(1130,425)(1138,426)(1147,427)(1155,427)(1164,428)(1173,428)(1181,429)(1190,429)(1198,429)(1207,429)(1215,429)(1224,429)(1232,429)(1241,429)(1250,429)(1258,429)(1267,429)(1275,428)(1284,428)(1292,428)(1301,428)(1309,428)(1318,428)(1327,428)(1335,428)(1344,429)(1352,429)(1361,429)(1369,429)(1378,429)(1387,429)(1395,429)(1404,429)(1412,429)(1421,429)(1429,429)(1438,429)(1446,429)(1455,428)(1464,428)(1472,428)(1481,428)(1489,428)(1498,428)
\thinlines \path(1498,428)(1506,427)(1515,427)(1523,427)(1532,427)(1541,426)(1549,426)(1558,426)(1566,426)(1575,426)(1583,426)(1592,426)(1600,426)(1609,426)(1618,427)(1626,427)(1635,427)(1643,427)(1652,427)(1660,425)
\thinlines \path(220,113)(226,113)(227,113)
\thinlines \path(239,113)(239,113)(240,113)(247,113)(253,113)(260,113)(263,113)
\thinlines \path(275,113)(275,113)(281,113)(287,114)(292,124)
\thinlines \path(295,134)(295,134)(297,155)
\thinlines \path(297,165)(297,165)(299,185)
\thinlines \path(300,196)(300,196)(301,203)(302,216)
\thinlines \path(302,227)(302,227)(303,247)
\thinlines \path(303,258)(303,258)(304,278)
\thinlines \path(304,289)(304,289)(305,309)
\thinlines \path(305,320)(305,320)(306,341)
\thinlines \path(306,351)(306,351)(307,372)
\thinlines \path(308,382)(308,382)(308,389)(308,403)
\thinlines \path(308,413)(308,413)(309,434)
\thinlines \path(309,444)(309,444)(310,465)
\thinlines \path(310,475)(310,475)(310,496)
\thinlines \path(311,506)(311,506)(311,527)
\thinlines \path(311,537)(311,537)(312,558)
\thinlines \path(312,568)(312,568)(313,589)
\thinlines \path(313,599)(313,599)(314,620)
\thinlines \path(314,630)(314,630)(314,651)
\thinlines \path(315,661)(315,661)(315,663)(315,682)
\thinlines \path(315,692)(315,692)(316,713)
\thinlines \path(316,723)(316,723)(317,744)
\thinlines \path(317,754)(317,754)(318,775)
\thinlines \path(318,785)(318,785)(319,806)
\thinlines \path(319,816)(319,816)(319,837)
\thinlines \path(320,847)(320,847)(320,868)
\thinlines \path(321,878)(321,878)(321,899)
\thinlines \path(322,909)(322,909)(323,930)
\thinlines \path(323,940)(323,940)(325,961)
\thinlines \path(325,971)(325,971)(326,992)
\thinlines \path(327,1002)(327,1002)(328,1023)
\thinlines \path(331,1018)(331,1018)(335,1010)(336,998)
\thinlines \path(337,988)(337,988)(338,967)
\thinlines \path(339,957)(339,957)(340,936)
\thinlines \path(341,926)(341,926)(342,912)(342,905)
\thinlines \path(343,895)(343,895)(344,874)
\thinlines \path(344,864)(344,864)(345,843)
\thinlines \path(346,833)(346,833)(347,812)
\thinlines \path(348,802)(348,802)(349,784)(349,781)
\thinlines \path(349,771)(349,771)(351,750)
\thinlines \path(351,740)(351,740)(352,719)
\thinlines \path(353,709)(353,709)(354,688)
\thinlines \path(355,678)(355,678)(355,668)(356,657)
\thinlines \path(357,647)(357,647)(358,626)
\thinlines \path(359,616)(359,616)(361,595)
\thinlines \path(361,585)(361,585)(362,575)(363,564)
\thinlines \path(364,554)(364,554)(366,533)
\thinlines \path(368,523)(368,523)(369,509)(370,502)
\thinlines \path(372,492)(372,492)(375,472)
\thinlines \path(377,461)(377,461)(382,441)
\thinlines \path(385,431)(385,431)(389,421)(395,412)
\thinlines \path(404,406)(404,406)(410,403)(416,402)(423,403)(428,403)
\thinlines \path(439,404)(439,404)(444,405)(450,408)(457,411)(462,413)
\thinlines \path(472,418)(472,418)(478,421)(484,426)(491,430)(492,430)
\thinlines \path(501,436)(501,436)(505,438)(512,443)(518,448)(521,449)
\thinlines \path(531,454)(531,454)(532,454)(539,458)(546,460)(552,461)(554,461)
\thinlines \path(566,461)(566,461)(566,461)(573,459)(580,457)(586,454)(588,453)
\thinlines \path(599,448)(599,448)(600,447)(607,444)(613,439)(618,436)
\thinlines \path(628,430)(628,430)(634,426)(641,421)(647,417)
\thinlines \path(656,410)(656,410)(661,407)(668,403)(675,400)(676,399)
\thinlines \path(687,393)(687,393)(688,392)(695,390)(702,388)(709,386)(709,386)
\thinlines \path(721,386)(721,386)(722,386)(729,387)(736,388)(743,390)(745,391)
\thinlines \path(755,396)(755,396)(756,396)(763,400)(770,404)(776,408)
\thinlines \path(785,413)(785,413)(790,416)(797,421)(804,425)(805,426)
\thinlines \path(815,431)(815,431)(817,432)(824,436)(831,440)(836,442)
\thinlines \path(846,447)(846,447)(851,449)(858,452)(865,454)(869,455)
\thinlines \path(880,458)(880,458)(885,459)(892,460)(899,460)(904,460)
\thinlines \path(916,459)(916,459)(919,458)(926,456)(933,454)(939,451)
\thinlines \path(949,446)(949,446)(953,444)(960,441)(967,437)(970,435)
\thinlines \path(981,430)(981,430)(987,427)(994,423)(1001,420)(1002,420)
\thinlines \path(1013,415)(1013,415)(1014,415)(1021,412)(1028,410)(1035,409)(1036,409)
\thinlines \path(1048,407)(1048,407)(1048,407)(1055,406)(1062,406)(1069,406)(1072,406)
\thinlines \path(1084,407)(1084,407)(1089,408)(1096,409)(1103,410)(1108,411)
\thinlines \path(1120,413)(1120,413)(1123,414)(1130,416)(1137,418)(1143,419)
\thinlines \path(1154,423)(1154,423)(1157,423)(1164,425)(1171,428)(1177,430)
\thinlines \path(1188,434)(1188,434)(1191,435)(1198,438)(1204,441)(1210,443)
\thinlines \path(1221,447)(1221,447)(1225,448)(1232,450)(1238,451)(1245,452)
\thinlines \path(1257,452)(1257,452)(1259,452)(1266,451)(1272,450)(1279,447)(1280,447)
\thinlines \path(1290,442)(1290,442)(1293,440)(1300,436)(1306,429)
\thinlines \path(221,113)(221,113)(223,113)(224,113)
\thinlines \path(237,113)(237,113)(240,113)(240,113)
\thinlines \path(253,113)(253,113)(256,113)
\thinlines \path(269,113)(269,113)(272,113)
\thinlines \path(284,115)(284,115)(286,116)
\thinlines \path(292,126)(292,126)(292,128)
\thinlines \path(293,140)(293,140)(294,142)
\thinlines \path(295,153)(295,153)(295,155)
\thinlines \path(297,167)(297,167)(297,169)
\thinlines \path(299,180)(299,180)(299,183)
\thinlines \path(300,194)(300,194)(300,196)
\thinlines \path(300,208)(300,208)(300,210)
\thinlines \path(301,221)(301,221)(301,223)
\thinlines \path(301,235)(301,235)(301,237)
\thinlines \path(302,248)(302,248)(302,251)
\thinlines \path(302,262)(302,262)(302,264)
\thinlines \path(303,276)(303,276)(303,278)
\thinlines \path(303,289)(303,289)(303,292)
\thinlines \path(304,303)(304,303)(304,305)
\thinlines \path(304,317)(304,317)(304,319)
\thinlines \path(305,330)(305,330)(305,333)
\thinlines \path(305,344)(305,344)(305,346)
\thinlines \path(306,358)(306,358)(306,360)
\thinlines \path(306,371)(306,371)(306,374)
\thinlines \path(307,385)(307,385)(307,387)
\thinlines \path(307,399)(307,399)(307,401)
\thinlines \path(308,412)(308,412)(308,414)
\thinlines \path(308,426)(308,426)(308,428)
\thinlines \path(308,440)(308,440)(309,442)
\thinlines \path(309,453)(309,453)(309,455)
\thinlines \path(309,467)(309,467)(309,469)
\thinlines \path(309,480)(309,480)(309,483)
\thinlines \path(310,494)(310,494)(310,496)
\thinlines \path(310,508)(310,508)(310,510)
\thinlines \path(310,521)(310,521)(310,524)
\thinlines \path(311,535)(311,535)(311,537)
\thinlines \path(311,549)(311,549)(311,551)
\thinlines \path(311,562)(311,562)(311,565)
\thinlines \path(312,576)(312,576)(312,578)
\thinlines \path(312,590)(312,590)(312,592)
\thinlines \path(312,603)(312,603)(312,606)
\thinlines \path(313,617)(313,617)(313,619)
\thinlines \path(313,631)(313,631)(313,633)
\thinlines \path(313,644)(313,644)(313,647)
\thinlines \path(314,658)(314,658)(314,660)
\thinlines \path(314,672)(314,672)(314,674)
\thinlines \path(314,685)(314,685)(314,687)
\thinlines \path(315,699)(315,699)(315,701)
\thinlines \path(315,713)(315,713)(315,715)
\thinlines \path(315,726)(315,726)(315,728)
\thinlines \path(316,740)(316,740)(316,742)
\thinlines \path(316,753)(316,753)(316,756)
\thinlines \path(316,767)(316,767)(316,769)
\thinlines \path(317,781)(317,781)(317,783)
\thinlines \path(317,794)(317,794)(317,797)
\thinlines \path(318,808)(318,808)(318,810)
\thinlines \path(318,822)(318,822)(318,824)
\thinlines \path(319,835)(319,835)(319,838)
\thinlines \path(319,849)(319,849)(319,851)
\thinlines \path(320,863)(320,863)(320,865)
\thinlines \path(320,876)(320,876)(320,879)
\thinlines \path(321,890)(321,890)(321,892)
\thinlines \path(321,904)(321,904)(321,906)
\thinlines \path(322,917)(322,917)(322,919)
\thinlines \path(322,931)(322,931)(322,933)
\thinlines \path(323,945)(323,945)(323,947)
\thinlines \path(323,958)(323,958)(323,960)
\thinlines \path(324,972)(324,972)(324,974)
\thinlines \path(324,985)(324,985)(324,988)
\thinlines \path(325,999)(325,999)(325,1001)
\thinlines \path(325,1013)(325,1013)(325,1015)
\thinlines \path(330,1025)(330,1025)(331,1027)
\thinlines \path(334,1025)(334,1025)(334,1022)
\thinlines \path(335,1011)(335,1011)(335,1009)
\thinlines \path(336,997)(336,997)(336,995)
\thinlines \path(337,984)(337,984)(337,981)
\thinlines \path(338,970)(338,970)(338,968)
\thinlines \path(339,956)(339,956)(339,954)
\thinlines \path(339,943)(339,943)(340,941)
\thinlines \path(340,929)(340,929)(340,927)
\thinlines \path(341,916)(341,916)(341,913)
\thinlines \path(342,902)(342,902)(342,900)
\thinlines \path(343,888)(343,888)(343,886)
\thinlines \path(343,875)(343,875)(344,872)
\thinlines \path(344,861)(344,861)(344,859)
\thinlines \path(345,847)(345,847)(345,845)
\thinlines \path(346,834)(346,834)(346,831)
\thinlines \path(346,820)(346,820)(346,818)
\thinlines \path(347,806)(347,806)(347,804)
\thinlines \path(348,793)(348,793)(348,791)
\thinlines \path(348,779)(348,779)(348,777)
\thinlines \path(349,766)(349,766)(349,763)
\thinlines \path(350,752)(350,752)(350,750)
\thinlines \path(350,738)(350,738)(351,736)
\thinlines \path(351,725)(351,725)(351,722)
\thinlines \path(352,711)(352,711)(352,709)
\thinlines \path(353,697)(353,697)(353,695)
\thinlines \path(354,684)(354,684)(354,681)
\thinlines \path(355,670)(355,670)(355,668)
\thinlines \path(355,656)(355,656)(356,654)
\thinlines \path(356,643)(356,643)(356,641)
\thinlines \path(357,629)(357,629)(357,627)
\thinlines \path(358,616)(358,616)(358,613)
\thinlines \path(359,602)(359,602)(359,600)
\thinlines \path(360,588)(360,588)(360,586)
\thinlines \path(361,575)(361,575)(361,572)
\thinlines \path(362,561)(362,561)(363,559)
\thinlines \path(364,547)(364,547)(364,545)
\thinlines \path(365,534)(365,534)(365,532)
\thinlines \path(366,520)(366,520)(366,518)
\thinlines \path(367,507)(367,507)(368,504)
\thinlines \path(369,493)(369,493)(370,491)
\thinlines \path(371,480)(371,480)(372,477)
\thinlines \path(374,466)(374,466)(374,464)
\thinlines \path(376,452)(376,452)(376,450)
\thinlines \path(379,439)(379,439)(379,437)
\thinlines \path(383,426)(383,426)(383,424)
\thinlines \path(388,413)(388,413)(389,411)
\thinlines \path(395,401)(395,401)(397,400)
\thinlines \path(409,394)(409,394)(411,394)(412,394)
\thinlines \path(425,395)(425,395)(427,396)
\thinlines \path(440,400)(440,400)(442,401)
\thinlines \path(454,406)(454,406)(456,408)
\thinlines \path(467,414)(467,414)(469,415)
\thinlines \path(479,423)(479,423)(481,424)
\thinlines \path(491,432)(491,432)(493,434)
\thinlines \path(503,441)(503,441)(505,443)
\thinlines \path(515,450)(515,450)(517,452)
\thinlines \path(528,458)(528,458)(530,460)
\thinlines \path(542,465)(542,465)(545,465)
\thinlines \path(558,466)(558,466)(560,466)
\thinlines \path(573,463)(573,463)(573,463)(576,462)
\thinlines \path(587,456)(587,456)(589,455)
\thinlines \path(600,448)(600,448)(602,447)
\thinlines \path(612,440)(612,440)(614,438)
\thinlines \path(624,430)(624,430)(625,430)(626,429)
\thinlines \path(636,421)(636,421)(638,420)
\thinlines \path(648,412)(648,412)(650,411)
\thinlines \path(661,404)(661,404)(663,402)
\thinlines \path(674,396)(674,396)(676,395)(676,395)
\thinlines \path(688,390)(688,390)(691,389)
\thinlines \path(703,386)(703,386)(706,386)
\thinlines \path(719,386)(719,386)(722,387)
\thinlines \path(735,390)(735,390)(736,390)(737,391)
\thinlines \path(749,397)(749,397)(751,398)
\thinlines \path(762,405)(762,405)(762,405)(764,406)
\thinlines \path(774,413)(774,413)(776,414)
\thinlines \path(787,421)(787,421)(787,422)(789,423)
\thinlines \path(800,429)(800,429)(802,430)
\thinlines \path(814,436)(814,436)(816,437)
\thinlines \path(828,442)(828,442)(830,443)(831,443)
\thinlines \path(843,447)(843,447)(846,448)
\thinlines \path(859,450)(859,450)(861,450)
\thinlines \path(875,451)(875,451)(877,451)
\thinlines \path(891,450)(891,450)(893,450)
\thinlines \path(906,447)(906,447)(907,447)(909,447)
\thinlines \path(922,444)(922,444)(924,443)(924,443)
\thinlines \path(937,439)(937,439)(940,439)
\thinlines \path(952,435)(952,435)(955,434)
\thinlines \path(967,431)(967,431)(970,430)
\thinlines \path(983,427)(983,427)(984,427)(985,426)
\thinlines \path(998,424)(998,424)(1001,423)
\thinlines \path(1014,422)(1014,422)(1017,421)
\thinlines \path(1030,420)(1030,420)(1033,420)
\thinlines \path(1046,420)(1046,420)(1049,420)
\thinlines \path(1062,419)(1062,419)(1065,419)
\thinlines \path(1078,420)(1078,420)(1078,420)(1081,420)
\thinlines \path(1094,421)(1094,421)(1096,421)(1097,421)
\thinlines \path(1110,421)(1110,421)(1113,421)(1113,421)
\thinlines \path(1126,422)(1126,422)(1129,422)
\thinlines \path(1142,423)(1142,423)(1145,424)
\thinlines \path(1158,424)(1158,424)(1161,425)
\thinlines \path(1174,426)(1174,426)(1177,426)
\thinlines \path(1190,427)(1190,427)(1193,427)
\thinlines \path(1206,429)(1206,429)(1207,429)(1209,429)
\thinlines \path(1222,430)(1222,430)(1224,431)(1224,431)
\thinlines \path(1238,432)(1238,432)(1240,432)
\thinlines \path(1254,433)(1254,433)(1256,433)
\thinlines \path(1270,434)(1270,434)(1272,434)
\thinlines \path(1286,434)(1286,434)(1288,434)
\thinlines \path(1302,433)(1302,433)(1304,433)
\thinlines \path(1318,432)(1318,432)(1318,432)(1320,432)
\thinlines \path(1334,430)(1334,430)(1335,430)(1336,430)
\thinlines \path(1349,429)(1349,429)(1352,428)
\thinlines \path(1365,427)(1365,427)(1368,427)
\thinlines \path(1381,425)(1381,425)(1384,425)
\thinlines \path(1397,424)(1397,424)(1400,424)
\thinlines \path(1413,423)(1413,423)(1416,423)
\thinlines \path(1429,423)(1429,423)(1429,422)(1432,422)
\thinlines \path(1445,422)(1445,422)(1446,422)(1448,422)
\thinlines \path(1461,423)(1461,423)(1464,423)(1464,423)
\thinlines \path(1477,423)(1477,423)(1480,423)
\thinlines \path(1493,424)(1493,424)(1496,424)
\thinlines \path(1509,425)(1509,425)(1512,425)
\thinlines \path(1525,427)(1525,427)(1528,427)
\thinlines \path(1541,429)(1541,429)(1543,429)
\thinlines \path(1557,432)(1557,432)(1558,432)(1559,432)
\thinlines \path(1572,435)(1572,435)(1575,435)(1575,435)
\thinlines \path(1588,437)(1588,437)(1591,437)
\thinlines \path(1604,438)(1604,438)(1607,438)
\thinlines \path(1620,438)(1620,438)(1623,438)
\thinlines \path(1636,436)(1636,436)(1638,436)
\thinlines \path(1651,432)(1651,432)(1652,432)(1653,431)
\end{picture}

%% file: grE.tex
\setlength{\unitlength}{0.240900pt}
\begin{picture}(1800,1080)(0,0)
\tenrm
\thicklines \path(220,113)(240,113)
\thicklines \path(1736,113)(1716,113)
\put(198,113){\makebox(0,0)[r]{0}}
\thicklines \path(220,428)(240,428)
\thicklines \path(1736,428)(1716,428)
\put(198,428){\makebox(0,0)[r]{1}}
\thicklines \path(220,742)(240,742)
\thicklines \path(1736,742)(1716,742)
\put(198,742){\makebox(0,0)[r]{2}}
\thicklines \path(220,1057)(240,1057)
\thicklines \path(1736,1057)(1716,1057)
\put(198,1057){\makebox(0,0)[r]{3}}
\thicklines \path(220,113)(220,133)
\thicklines \path(220,1057)(220,1037)
\put(220,68){\makebox(0,0){2}}
\thicklines \path(496,113)(496,133)
\thicklines \path(496,1057)(496,1037)
\put(496,68){\makebox(0,0){4}}
\thicklines \path(771,113)(771,133)
\thicklines \path(771,1057)(771,1037)
\put(771,68){\makebox(0,0){6}}
\thicklines \path(1047,113)(1047,133)
\thicklines \path(1047,1057)(1047,1037)
\put(1047,68){\makebox(0,0){8}}
\thicklines \path(1323,113)(1323,133)
\thicklines \path(1323,1057)(1323,1037)
\put(1323,68){\makebox(0,0){10}}
\thicklines \path(1598,113)(1598,133)
\thicklines \path(1598,1057)(1598,1037)
\put(1598,68){\makebox(0,0){12}}
\thicklines \path(220,113)(1736,113)(1736,1057)(220,1057)(220,113)
\put(0,585){\makebox(0,0)[l]{\shortstack{g$_{OO}$(r)}}}
\put(978,23){\makebox(0,0){r ({\AA})}}
\thinlines \path(220,113)(223,113)(231,113)(240,113)(248,113)(257,113)(265,113)(274,113)(282,113)(291,120)(300,185)(308,414)(317,762)(325,1003)(334,1025)(342,898)(351,734)(359,599)(368,508)(377,452)(385,423)(394,406)(402,400)(411,398)(419,398)(428,399)(436,402)(445,404)(454,408)(462,413)(471,419)(479,424)(488,430)(496,437)(505,443)(514,449)(522,455)(531,459)(539,462)(548,464)(556,465)(565,464)(573,461)(582,458)(591,453)(599,448)(608,442)(616,435)(625,429)(633,423)(642,417)
\thinlines \path(642,417)(650,410)(659,404)(668,399)(676,394)(685,391)(693,388)(702,385)(710,384)(719,385)(727,387)(736,389)(745,393)(753,399)(762,404)(770,409)(779,415)(787,421)(796,427)(805,432)(813,437)(822,441)(830,445)(839,448)(847,450)(856,452)(864,453)(873,454)(882,453)(890,452)(899,451)(907,449)(916,447)(924,444)(933,441)(941,438)(950,435)(959,432)(967,429)(976,427)(984,425)(993,423)(1001,421)(1010,420)(1018,419)(1027,419)(1036,418)(1044,418)(1053,418)(1061,419)(1070,419)
\thinlines \path(1070,419)(1078,420)(1087,420)(1096,421)(1104,422)(1113,423)(1121,424)(1130,425)(1138,426)(1147,427)(1155,428)(1164,428)(1173,428)(1181,429)(1190,429)(1198,429)(1207,429)(1215,429)(1224,429)(1232,429)(1241,429)(1250,429)(1258,429)(1267,429)(1275,428)(1284,428)(1292,428)(1301,428)(1309,428)(1318,428)(1327,428)(1335,428)(1344,428)(1352,429)(1361,429)(1369,429)(1378,429)(1387,429)(1395,429)(1404,429)(1412,429)(1421,429)(1429,429)(1438,429)(1446,429)(1455,428)(1464,428)(1472,428)(1481,428)(1489,427)(1498,427)
\thinlines \path(1498,427)(1506,427)(1515,427)(1523,427)(1532,427)(1541,426)(1549,426)(1558,426)(1566,426)(1575,426)(1583,426)(1592,426)(1600,426)(1609,426)(1618,427)(1626,427)(1635,427)(1643,427)(1652,427)(1660,425)
\thinlines \path(220,113)(226,113)
\thinlines \path(238,113)(238,113)(240,113)(247,113)(253,113)(260,113)(262,113)
\thinlines \path(275,113)(275,113)(281,113)(287,115)(290,124)
\thinlines \path(293,134)(293,134)(294,137)(295,155)
\thinlines \path(296,165)(296,165)(298,186)
\thinlines \path(298,196)(298,196)(300,216)
\thinlines \path(301,227)(301,227)(301,233)(301,247)
\thinlines \path(302,258)(302,258)(302,278)
\thinlines \path(303,289)(303,289)(303,309)
\thinlines \path(304,320)(304,320)(304,340)
\thinlines \path(305,350)(305,350)(305,371)
\thinlines \path(306,381)(306,381)(306,402)
\thinlines \path(307,412)(307,412)(307,433)
\thinlines \path(307,443)(307,443)(308,454)(308,464)
\thinlines \path(308,474)(308,474)(309,495)
\thinlines \path(309,505)(309,505)(309,526)
\thinlines \path(310,536)(310,536)(310,557)
\thinlines \path(310,567)(310,567)(311,588)
\thinlines \path(311,598)(311,598)(312,619)
\thinlines \path(312,629)(312,629)(312,649)
\thinlines \path(313,660)(313,660)(313,680)
\thinlines \path(313,691)(313,691)(314,711)
\thinlines \path(314,722)(314,722)(314,742)
\thinlines \path(315,753)(315,753)(315,773)
\thinlines \path(316,784)(316,784)(316,804)
\thinlines \path(316,814)(316,814)(317,835)
\thinlines \path(317,845)(317,845)(318,866)
\thinlines \path(318,876)(318,876)(319,897)
\thinlines \path(319,907)(319,907)(320,928)
\thinlines \path(320,938)(320,938)(321,959)
\thinlines \path(321,969)(321,969)(321,988)(322,990)
\thinlines \path(322,1000)(322,1000)(324,1021)
\thinlines \path(325,1031)(325,1031)(326,1052)
\thinlines \path(331,1055)(331,1055)(333,1035)
\thinlines \path(335,1025)(335,1025)(335,1023)(336,1004)
\thinlines \path(337,994)(337,994)(338,973)
\thinlines \path(338,963)(338,963)(339,942)
\thinlines \path(340,932)(340,932)(341,911)
\thinlines \path(341,901)(341,901)(342,895)(342,880)
\thinlines \path(343,870)(343,870)(344,850)
\thinlines \path(344,839)(344,839)(345,819)
\thinlines \path(346,808)(346,808)(347,788)
\thinlines \path(347,777)(347,777)(348,757)
\thinlines \path(349,746)(349,746)(350,726)
\thinlines \path(351,716)(351,716)(352,695)
\thinlines \path(352,685)(352,685)(353,664)
\thinlines \path(354,654)(354,654)(355,633)
\thinlines \path(356,623)(356,623)(357,602)
\thinlines \path(358,592)(358,592)(359,571)
\thinlines \path(360,561)(360,561)(362,540)
\thinlines \path(363,530)(363,530)(365,510)
\thinlines \path(366,499)(366,499)(368,479)
\thinlines \path(369,469)(369,469)(372,448)
\thinlines \path(374,438)(374,438)(376,429)(379,418)
\thinlines \path(382,408)(382,408)(383,404)(389,390)(391,389)
\thinlines \path(400,382)(400,382)(403,380)(410,379)(416,381)(422,383)
\thinlines \path(433,388)(433,388)(437,390)(444,394)(450,399)(452,400)
\thinlines \path(461,407)(461,407)(464,409)(471,414)(478,421)(479,422)
\thinlines \path(486,430)(486,430)(491,434)(498,441)(502,445)
\thinlines \path(510,453)(510,453)(512,454)(518,459)(525,465)(528,467)
\thinlines \path(538,472)(538,472)(539,472)(546,475)(552,476)(559,476)(562,475)
\thinlines \path(573,472)(573,472)(580,469)(586,465)(593,460)
\thinlines \path(601,453)(601,453)(607,449)(613,442)(619,438)
\thinlines \path(627,431)(627,431)(627,431)(634,425)(641,419)(644,416)
\thinlines \path(653,409)(653,409)(654,408)(661,403)(668,398)(671,396)
\thinlines \path(681,390)(681,390)(681,389)(688,386)(695,384)(702,382)(703,382)
\thinlines \path(715,380)(715,380)(715,380)(722,380)(729,382)(736,385)(738,386)
\thinlines \path(748,391)(748,391)(749,392)(756,396)(763,400)(767,404)
\thinlines \path(777,410)(777,410)(783,416)(790,420)(795,424)
\thinlines \path(805,430)(805,430)(810,434)(817,438)(824,441)(825,442)
\thinlines \path(835,446)(835,446)(838,447)(844,450)(851,451)(858,452)(858,452)
\thinlines \path(870,454)(870,454)(872,454)(878,454)(885,454)(892,453)(894,453)
\thinlines \path(906,450)(906,450)(912,448)(919,446)(926,444)(929,443)
\thinlines \path(940,439)(940,439)(946,436)(953,433)(960,432)(962,431)
\thinlines \path(974,428)(974,428)(980,426)(987,424)(994,423)(997,422)
\thinlines \path(1009,421)(1009,421)(1014,420)(1021,420)(1028,419)(1033,419)
\thinlines \path(1045,419)(1045,419)(1048,419)(1055,419)(1062,419)(1069,419)(1069,419)
\thinlines \path(1081,421)(1081,421)(1082,421)(1089,421)(1096,422)(1103,423)(1105,423)
\thinlines \path(1117,424)(1117,424)(1123,425)(1130,426)(1137,426)(1141,426)
\thinlines \path(1153,427)(1153,427)(1157,427)(1164,428)(1171,428)(1177,428)(1178,428)
\thinlines \path(1190,428)(1190,428)(1191,428)(1198,428)(1204,429)(1211,428)(1214,428)
\thinlines \path(1226,428)(1226,428)(1232,428)(1238,428)(1245,428)(1250,428)
\thinlines \path(1262,428)(1262,428)(1266,428)(1272,427)(1279,427)(1286,427)(1286,427)
\thinlines \path(1298,426)(1298,426)(1300,426)(1306,424)
\thinlines \path(222,113)(222,113)(223,113)(225,113)
\thinlines \path(238,113)(238,113)(240,113)(241,113)
\thinlines \path(254,113)(254,113)(257,113)(257,113)
\thinlines \path(270,113)(270,113)(273,113)
\thinlines \path(285,116)(285,116)(287,118)
\thinlines \path(292,128)(292,128)(292,130)
\thinlines \path(293,141)(293,141)(293,144)
\thinlines \path(295,155)(295,155)(295,157)
\thinlines \path(296,169)(296,169)(296,171)
\thinlines \path(298,182)(298,182)(298,185)
\thinlines \path(299,196)(299,196)(299,198)
\thinlines \path(300,210)(300,210)(300,212)
\thinlines \path(300,223)(300,223)(300,226)
\thinlines \path(301,237)(301,237)(301,239)
\thinlines \path(301,251)(301,251)(301,253)
\thinlines \path(302,264)(302,264)(302,267)
\thinlines \path(302,278)(302,278)(302,280)
\thinlines \path(303,292)(303,292)(303,294)
\thinlines \path(303,306)(303,306)(303,308)
\thinlines \path(303,319)(303,319)(304,322)
\thinlines \path(304,333)(304,333)(304,335)
\thinlines \path(304,347)(304,347)(304,349)
\thinlines \path(305,360)(305,360)(305,363)
\thinlines \path(305,374)(305,374)(305,376)
\thinlines \path(306,388)(306,388)(306,390)
\thinlines \path(306,402)(306,402)(306,404)
\thinlines \path(307,415)(307,415)(307,418)
\thinlines \path(307,429)(307,429)(307,431)
\thinlines \path(308,443)(308,443)(308,445)
\thinlines \path(308,456)(308,456)(308,459)
\thinlines \path(308,470)(308,470)(308,472)
\thinlines \path(309,484)(309,484)(309,486)
\thinlines \path(309,497)(309,497)(309,500)
\thinlines \path(309,511)(309,511)(309,513)
\thinlines \path(310,525)(310,525)(310,527)
\thinlines \path(310,539)(310,539)(310,541)
\thinlines \path(310,552)(310,552)(310,555)
\thinlines \path(311,566)(311,566)(311,568)
\thinlines \path(311,580)(311,580)(311,582)
\thinlines \path(311,593)(311,593)(311,596)
\thinlines \path(312,607)(312,607)(312,609)
\thinlines \path(312,621)(312,621)(312,623)
\thinlines \path(312,635)(312,635)(312,637)
\thinlines \path(313,648)(313,648)(313,651)
\thinlines \path(313,662)(313,662)(313,664)
\thinlines \path(313,676)(313,676)(313,678)
\thinlines \path(314,689)(314,689)(314,692)
\thinlines \path(314,703)(314,703)(314,705)
\thinlines \path(314,717)(314,717)(314,719)
\thinlines \path(315,730)(315,730)(315,733)
\thinlines \path(315,744)(315,744)(315,746)
\thinlines \path(315,758)(315,758)(315,760)
\thinlines \path(315,772)(315,772)(316,774)
\thinlines \path(316,785)(316,785)(316,788)
\thinlines \path(316,799)(316,799)(316,801)
\thinlines \path(316,813)(316,813)(317,815)
\thinlines \path(317,826)(317,826)(317,829)
\thinlines \path(317,840)(317,840)(317,842)
\thinlines \path(318,854)(318,854)(318,856)
\thinlines \path(318,868)(318,868)(319,870)
\thinlines \path(319,881)(319,881)(319,884)
\thinlines \path(319,895)(319,895)(320,897)
\thinlines \path(320,909)(320,909)(320,911)
\thinlines \path(321,922)(321,922)(321,925)
\thinlines \path(321,936)(321,936)(321,938)
\thinlines \path(322,950)(322,950)(322,952)
\thinlines \path(322,963)(322,963)(322,966)
\thinlines \path(323,977)(323,977)(323,979)
\thinlines \path(323,991)(323,991)(323,993)
\thinlines \path(324,1005)(324,1005)(324,1007)
\thinlines \path(324,1018)(324,1018)(324,1021)
\thinlines \path(325,1032)(325,1032)(325,1034)
\thinlines \path(326,1044)(326,1044)(327,1042)
\thinlines \path(334,1033)(334,1033)(334,1031)
\thinlines \path(335,1019)(335,1019)(335,1017)
\thinlines \path(335,1005)(335,1005)(336,1003)
\thinlines \path(336,992)(336,992)(336,989)
\thinlines \path(337,978)(337,978)(337,976)
\thinlines \path(338,964)(338,964)(338,962)
\thinlines \path(339,951)(339,951)(339,948)
\thinlines \path(339,937)(339,937)(339,935)
\thinlines \path(340,923)(340,923)(340,921)
\thinlines \path(341,910)(341,910)(341,907)
\thinlines \path(342,896)(342,896)(342,894)
\thinlines \path(342,882)(342,882)(343,880)
\thinlines \path(343,869)(343,869)(343,866)
\thinlines \path(344,855)(344,855)(344,853)
\thinlines \path(344,841)(344,841)(345,839)
\thinlines \path(345,827)(345,827)(345,825)
\thinlines \path(346,814)(346,814)(346,811)
\thinlines \path(346,800)(346,800)(347,798)
\thinlines \path(347,786)(347,786)(347,784)
\thinlines \path(348,773)(348,773)(348,770)
\thinlines \path(348,759)(348,759)(349,757)
\thinlines \path(349,745)(349,745)(349,743)
\thinlines \path(350,732)(350,732)(350,729)
\thinlines \path(350,718)(350,718)(351,716)
\thinlines \path(351,704)(351,704)(351,702)
\thinlines \path(352,690)(352,690)(352,688)
\thinlines \path(353,677)(353,677)(353,674)
\thinlines \path(354,663)(354,663)(354,661)
\thinlines \path(355,649)(355,649)(355,647)
\thinlines \path(356,636)(356,636)(356,633)
\thinlines \path(356,622)(356,622)(357,620)
\thinlines \path(357,608)(357,608)(357,606)
\thinlines \path(358,595)(358,595)(358,592)
\thinlines \path(359,581)(359,581)(359,579)
\thinlines \path(360,567)(360,567)(360,565)
\thinlines \path(361,554)(361,554)(362,551)
\thinlines \path(363,540)(363,540)(363,538)
\thinlines \path(364,526)(364,526)(364,524)
\thinlines \path(365,513)(365,513)(365,510)
\thinlines \path(366,499)(366,499)(367,497)
\thinlines \path(368,485)(368,485)(368,483)
\thinlines \path(370,472)(370,472)(370,469)
\thinlines \path(372,458)(372,458)(372,456)
\thinlines \path(374,445)(374,445)(375,442)
\thinlines \path(377,431)(377,431)(377,431)(377,429)
\thinlines \path(380,418)(380,418)(381,415)
\thinlines \path(384,404)(384,404)(385,402)
\thinlines \path(392,392)(392,392)(393,390)
\thinlines \path(404,384)(404,384)(407,384)
\thinlines \path(420,386)(420,386)(422,386)
\thinlines \path(434,391)(434,391)(436,392)(437,392)
\thinlines \path(447,399)(447,399)(450,400)
\thinlines \path(460,407)(460,407)(462,409)
\thinlines \path(472,416)(472,416)(474,418)
\thinlines \path(483,426)(483,426)(485,428)
\thinlines \path(494,436)(494,436)(496,438)
\thinlines \path(505,447)(505,447)(505,447)(507,448)
\thinlines \path(516,456)(516,456)(518,458)
\thinlines \path(528,465)(528,465)(531,467)
\thinlines \path(543,472)(543,472)(545,472)
\thinlines \path(558,473)(558,473)(561,473)
\thinlines \path(574,469)(574,469)(576,468)
\thinlines \path(587,462)(587,462)(589,461)
\thinlines \path(599,453)(599,453)(601,451)
\thinlines \path(611,444)(611,444)(613,442)
\thinlines \path(623,434)(623,434)(624,432)
\thinlines \path(634,424)(634,424)(636,423)
\thinlines \path(646,415)(646,415)(647,413)
\thinlines \path(657,405)(657,405)(659,404)(659,404)
\thinlines \path(670,397)(670,397)(672,396)
\thinlines \path(683,389)(683,389)(685,389)(686,388)
\thinlines \path(698,384)(698,384)(700,383)
\thinlines \path(714,381)(714,381)(716,381)
\thinlines \path(729,383)(729,383)(732,384)
\thinlines \path(744,389)(744,389)(745,390)(746,391)
\thinlines \path(757,397)(757,397)(759,399)
\thinlines \path(769,406)(769,406)(770,407)(771,407)
\thinlines \path(782,415)(782,415)(784,416)
\thinlines \path(794,424)(794,424)(796,425)(796,425)
\thinlines \path(807,432)(807,432)(809,433)
\thinlines \path(820,440)(820,440)(822,440)(822,441)
\thinlines \path(834,446)(834,446)(837,447)
\thinlines \path(849,451)(849,451)(852,452)
\thinlines \path(865,454)(865,454)(868,454)
\thinlines \path(881,454)(881,454)(882,454)(884,454)
\thinlines \path(897,452)(897,452)(899,452)(900,452)
\thinlines \path(912,448)(912,448)(915,447)
\thinlines \path(927,443)(927,443)(930,442)
\thinlines \path(942,438)(942,438)(945,437)
\thinlines \path(957,433)(957,433)(959,432)(960,432)
\thinlines \path(972,428)(972,428)(975,427)
\thinlines \path(988,424)(988,424)(990,424)
\thinlines \path(1003,421)(1003,421)(1006,421)
\thinlines \path(1019,419)(1019,419)(1022,419)
\thinlines \path(1035,418)(1035,418)(1036,418)(1038,418)
\thinlines \path(1051,418)(1051,418)(1053,419)(1054,419)
\thinlines \path(1067,419)(1067,419)(1070,420)(1070,420)
\thinlines \path(1083,420)(1083,420)(1086,421)
\thinlines \path(1099,422)(1099,422)(1102,423)
\thinlines \path(1115,424)(1115,424)(1118,424)
\thinlines \path(1131,425)(1131,425)(1134,426)
\thinlines \path(1147,427)(1147,427)(1150,427)
\thinlines \path(1163,428)(1163,428)(1164,428)(1166,428)
\thinlines \path(1179,429)(1179,429)(1181,429)(1182,429)
\thinlines \path(1195,429)(1195,429)(1198,429)
\thinlines \path(1212,429)(1212,429)(1214,429)
\thinlines \path(1228,429)(1228,429)(1230,428)
\thinlines \path(1244,428)(1244,428)(1246,428)
\thinlines \path(1260,428)(1260,428)(1262,428)
\thinlines \path(1276,428)(1276,428)(1279,428)
\thinlines \path(1292,427)(1292,427)(1292,427)(1295,427)
\thinlines \path(1308,428)(1308,428)(1309,428)(1311,428)
\thinlines \path(1324,428)(1324,428)(1327,428)(1327,428)
\thinlines \path(1340,428)(1340,428)(1343,428)
\thinlines \path(1356,428)(1356,428)(1359,429)
\thinlines \path(1372,429)(1372,429)(1375,429)
\thinlines \path(1388,429)(1388,429)(1391,429)
\thinlines \path(1405,430)(1405,430)(1407,430)
\thinlines \path(1421,429)(1421,429)(1421,429)(1423,429)
\thinlines \path(1437,429)(1437,429)(1438,429)(1439,429)
\thinlines \path(1453,429)(1453,429)(1455,429)(1455,429)
\thinlines \path(1469,428)(1469,428)(1472,428)
\thinlines \path(1485,428)(1485,428)(1488,428)
\thinlines \path(1501,427)(1501,427)(1504,427)
\thinlines \path(1517,427)(1517,427)(1520,427)
\thinlines \path(1533,426)(1533,426)(1536,426)
\thinlines \path(1549,426)(1549,426)(1552,426)
\thinlines \path(1565,426)(1565,426)(1566,426)(1568,426)
\thinlines \path(1581,426)(1581,426)(1583,426)(1584,426)
\thinlines \path(1597,426)(1597,426)(1600,426)
\thinlines \path(1614,426)(1614,426)(1616,426)
\thinlines \path(1630,427)(1630,427)(1632,427)
\thinlines \path(1646,427)(1646,427)(1648,426)
\end{picture}

%% file: ewald.bbl
\begin{thebibliography}{10}

\bibitem{Watts}
Watts, R.~O., {\em Mol. Phys.}, {\bf4}, 1069--1083, (1974).

\bibitem{PangaliRaoBerne2}
Pangali, C., Rao, M.,  and Berne, B.~J., {\em Mol. Phys.}, {\bf40}, 661--680,
  (1980).

\bibitem{BrooksPettittKarplus}
{Brooks, III}, C.~L., Pettitt, B.~M.,  and Karplus, M., {\em J. Chem. Phys.},
  {\bf83}, 5897--5908, (1985).

\bibitem{Prevost}
Prevost, M., van Belle, D., Lippens, G.,  and Wodak, S., {\em Mol. Phys.},
  {\bf71}, 587--603, (1990).

\bibitem{Hunenberger}
{H\"{u}nenberger}, P.~H. and van Gunsteren, W.~F., {\em J. Chem. Phys.},
  {\bf108}, 6117--6134, (1998).

\bibitem{NewBerne}
New, M.~H. and Berne, B.~J., {\em J. Am. Chem. Soc.}, {\bf117}, 7172--7179,
  (1995).

\bibitem{LoncharichBrooks}
Loncharich, R.~J. and Brooks, B.~R., {\em Proteins: Struct., Funct., Genet.},
  {\bf6}, 32--45, (1989).

\bibitem{SmithPettitt}
Smith, P.~E. and Pettitt, B.~M., {\em J. Chem. Phys.}, {\bf95}, 8430--8441,
  (1991).

\bibitem{SchreiberSteinhauser}
Schreiber, H. and Steinhauser, O., {\em Biochemistry}, {\bf31}, 5856--5860,
  (1992).

\bibitem{Darden}
York, D.~M., Darden, T.~A.,  and Pedersen, L.~G., {\em J. Chem. Phys.},
  {\bf99}, 8345--8348, (1993).

\bibitem{SteinbachBrooks}
Steinbach, P.~J. and Brooks, B.~R., {\em J. Comput. Chem.}, {\bf15}, 667--683,
  (1994).

\bibitem{Greengard}
Greengard, L. and Rokhlin, V., {\em J. Comput. Phys.}, {\bf73}, 325--348,
  (1987).

\bibitem{EastwoodHockney}
Eastwood, J.~W. and Hockney, R.~W., {\em J. Comp. Phys}, {\bf16}, 342--359,
  (1974).

\bibitem{Darden2}
Darden, T., York, D.,  and Pedersen, L., {\em J. Chem. Phys.}, {\bf98},
  10089--10092, (1993).

\bibitem{Allen}
Allen, M.~P. and Tildesley, D.~J., {\em Computer Simulation of Liquids}.
\newblock Oxford University Press, Oxford, 1987.

\bibitem{BarkerWatts}
Barker, J.~A. and Watts, R.~O., {\em Mol. Phys.}, {\bf26}, 789--792, (1973).

\bibitem{Barnes}
Barnes, P., Finney, J.~L., Nicholas, J.~D.,  and Quinn, J.~E., {\em Nature},
  {\bf282}, 459--464, (1979).

\bibitem{Ahlstrom}
{Ahlstr\"{o}m}, P., Wallqvist, A., {Engstr\"{o}m}, S.,  and {J\"{o}nsson}, B.,
  {\em Mol. Phys.}, {\bf68}, 563--581, (1989).

\bibitem{KozackJordon}
Kozack, R.~E. and Jordon, P.~C., {\em J. Chem. Phys.}, {\bf96}, 3120--3130,
  (1992).

\bibitem{vanBelle}
van Belle, D. and Wodak, S.~J., {\em J. Am. Chem. Soc.}, {\bf115}, 647--652,
  (1993).

\bibitem{BernardoLevy}
Bernardo, D.~N., Ding, Y., Krogh-Jespersen, K.,  and Levy, R.~M., {\em J. Phys.
  Chem.}, {\bf98}, 4180--4187, (1994).

\bibitem{Zhu}
Zhu, S.-B. and Wong, C.~F., {\em J. Phys. Chem.}, {\bf98}, 4695--4701, (1994).

\bibitem{CaldwellKollman}
Caldwell, J.~W. and Kollman, P.~A., {\em J. Phys. Chem.}, {\bf99}, 6208--6219,
  (1995).

\bibitem{Soetens}
Soetens, J.-C. and Millot, C., {\em Chem. Phys. Lett.}, {\bf235}, 22--30,
  (1995).

\bibitem{ChialvoCummings}
Chialvo, A.~A. and Cummings, P.~T., {\em J. Chem. Phys.}, {\bf105}, 8274--8281,
  (1996).

\bibitem{DangChang}
Dang, L.~X. and Chang, T.~M., {\em J. Chem. Phys.}, {\bf106}, 8149--8159,
  (1997).

\bibitem{Siepmann}
Martin, M.~G., Chen, B.,  and Siepmann, J.~I., {\em J. Chem. Phys.}, {\bf108},
  3383--3385, (1998).

\bibitem{RickStuartBerne}
Rick, S.~W., Stuart, S.~J.,  and Berne, B.~J., {\em J. Chem. Phys.}, {\bf101},
  6141--6156, (1994).

\bibitem{Linse}
Linse, P. and Andersen, H.~C., {\em J. Chem. Phys.}, {\bf85}, 3027--3041,
  (1986).

\bibitem{JorgensenKlein}
Jorgensen, W.~L., Chandrasekhar, J., Madura, J.~D., Impey, R.~W.,  and Klein,
  M.~L., {\em J. Chem. Phys.}, {\bf79}, 926--935, (1983).

\bibitem{Nose}
{Nos\'{e}}, S., {\em Mol. Phys.}, {\bf52}, 255--268, (1984).

\bibitem{Hoover}
Hoover, W.~G., {\em Phys. Rev. A}, {\bf31}, 1695--1697, (1985).

\bibitem{RyckaertCiccottiBerendsen}
Ryckaert, J.~P., Ciccotti, G.,  and Berendsen, H. J.~C., {\em J. Comput.
  Phys.}, {\bf23}, 327--341, (1977).

\bibitem{JorgensenMadura}
Jorgensen, W.~L. and Madura, J.~D., {\em Mol Phys}, {\bf56}, 1381--1392,
  (1985).

\bibitem{SmithDang}
Smith, D.~E. and Dang, L.~X., {\em J. Chem. Phys.}, {\bf100}, 3757--3766,
  (1993).

\bibitem{Dang}
Dang, L.~X., {\em J. Chem. Phys.}, {\bf97}, 2659--2660, (1992).

\bibitem{Hummer}
Hummer, G., Gr{\o}nbech-Jensen, N.,  and Neumann, M., {\em J. Chem. Phys.},
  {\bf109}, 2791--2797, (1998).

\bibitem{Krynicki}
Krynicki, K., Green, C.~D.,  and Sawyer, D.~W., {\em Discuss. Faraday Soc.},
  {\bf66}, 199--208, (1978).

\bibitem{Jonas}
Jonas, J., DeFries, T.,  and Wilber, D.~J., {\em J. Chem. Phys.}, {\bf65},
  582--588, (1976).

\bibitem{Impey}
Impey, R.~W., Madden, P.~A.,  and McDonald, I.~R., {\em Mol. Phys.}, {\bf46},
  513--539, (1982).

\bibitem{FrenkelSmit}
Frenkel, D. and Smit, B., {\em Understanding Molecular Simulation: from
  Algorithms to Applications}.
\newblock {Academic Press, San Diego}, 1996.

\end{thebibliography}
